\documentclass{aa}

\usepackage{graphicx}
\usepackage{txfonts}
\usepackage{xcolor}
\usepackage{lineno}
\usepackage[normalem]{ulem}

\newcommand{\revise}[1]{#1}
\begin{document}

   \title{Giant planet  formation via pebble accretion across different stellar masses}


    \author{S. Shibata
            \inst{1,2}
            \thanks{E-mail: s.shibata423@gmail.com}
            \and
            R. Helled\inst{2}
    }
    
    
    \institute{Department of Earth, Environmental and Planetary Sciences, 6100 Main MS 126, Rice            University, Houston, TX 77005, USA\\
         \and
            Department of Astrophysics, University of Zurich, Winterthurerstrasse 190, CH-8057 Zurich, Switzerland
            }

   \date{Received June 1, 2025; accepted July 30, 2025}

 
  \abstract
   {}
   {
   The occurrence rate of cold Jupiters, giant planets orbiting in the outer orbital region ($\gtrsim1$ au), was found to depend on stellar mass. The formation environment in the protoplanetary disks, which depends on the mass of the host star, regulates core formation and the subsequent gas accretion. In this study, we simulate giant planet formation via pebble accretion accounting for various  stellar masses, core formation times, disk turbulent viscosities, and grain opacities.  
    }
   {
   We use a self-consistent formation model that calculates the solid accretion rate and gas accretion rate of growing protoplanets. We investigate how the planetary formation, in particular, the contraction of the envelope, and the formation timescale change under different conditions.  
   }
   {
    We find that to reproduce the observed occurrence rate of cold Jupiters, giant planets must undergo slow envelope contraction after they reach  pebble isolation, which lasts for several Myrs. Such a slow contraction phase can be achieved when the grain opacity is assumed to be as high as that of the interstellar medium (ISM). If the grain opacity is smaller than the ISM opacity by a factor of ten or more, the growing protoplanets reach crossover mass within 3 Myrs and form too many cold Jupiters around stars of $\gtrsim0.4M_\odot$. Protoplanets around low-mass stars $<0.4M_\odot$ take $\gtrsim 10$ Myrs to reach crossover mass also with low grain opacity.
    \revise{If} the grain opacity in the planetary envelope \revise{is} much lower than that of ISM, \revise{other mechanisms}, such as atmospheric recycling or \revise{planetesimal accretion}, is required for cold Jupiter formation. We next explore how the deposition of the accreted heavy elements to the planetary envelope changes the formation timescale. Our model suggests that the formation timescale could be longer due to heavy-element enrichment, resulting from the lower core mass at pebble isolation. We conclude that \revise{the details of the formation processes have} a significant effect on the planetary growth and therefore, the formation of gaseous planets. 
  }
   {
   }


   \maketitle
%
\section{Introduction}

Cold Jupiters are massive gaseous planets, usually categorized as those heavier than Neptune ($\gtrsim 30 M_\oplus$) that orbit their host star beyond the inferred ice lines ($\gtrsim 1$ au). The occurrence rate of cold Jupiters is usually constrained with RV observations \citep{Johnson+2009, Montet+2014, Rosenthal+2021, Fulton+2021, Hirsch+2021, Ribas+2022, Bonomo+2023, Pass+2023}. The observational data suggest that cold Jupiters are more common around more massive stars with higher metallicity. While the occurrence rate of cold Jupiters is constrained around $20 \%$ around G-type stars, it decreases lower than $10\%$ around M- and K-type stars \citep{Fulton+2021}.
The occurrence rate is rather low (less than $1.5\%$) around the low-mass ($0.1-0.3 M_\odot$) M dwarfs \citep{Pass+2023}, but two confirmed planets, LHS 252 b \citep{Morales+2019} and  GJ 83.1 b \citep{Feng+2020, Quirrenbach+2022}, and several candidates are detected there.


This result has been interpreted as a confirmation of the core accretion model \citep[e.g.][]{Mizuno1980} since the solid material required for forming heavy-element cores is more abundant around those stars. In the core formation scenario, planetesimal and/or pebble accretion form a massive solid core ($\sim10M_\oplus$). The solid core accretes surrounding disk gas and forms a gaseous envelope around the core. Once the protoplanet reaches the crossover mass, where the envelope mass equals the solid core mass, the protoplanet starts a rapid gas accretion \citep{Pollack+1996, Ikoma+2000}. The crossover mass is usually considered the starting point of runaway gas accretion. To form gas giant planets, protoplanets must reach the crossover mass before the disk dissipation. In the core accretion model, the occurrence rate of giant planets depends on the timescales of core formation and the subsequent gas accretion. Besides the core accretion model, the gravitational instability scenario \citep{Boss2006, Boss2019} is also suggested to form giant planets around low-mass M dwarfs \citep{Morales+2019}. However, the relation between the occurrence rate and the stellar metallicity is not explained in this formation mechanism \citep{Mercer+2020}. Therefore, the core accretion model would be the primary formation mechanism of cold Jupiters, while some of the giant planets around lower mass stars might be formed via gravitational instability.

Various formation models investigated how the occurrence rate of giant planets changes with the assumed stellar properties. The planetesimal accretion scenario is the classical formation model for a massive solid core \citep{Inaba+2003, Kobayashi+2021, Kobayashi+2023}. Population synthesis models based on the planetesimal accretion scenario \citep{Ida+2005, Alibert+2011, Miguel+2019, Burn+2021} show that the formation rate of giant planets increases with stellar mass, which is consistent with the observations. However, the planetesimal accretion scenario has difficulty in the formation of giant planets around M dwarfs since current planetesimal accretion model failed to form a massive solid core \citep{Miguel+2019, Schlecker+2022}. 

In the pebble accretion scenario, protoplanets grow via accretion of small millimeter-to-centimeter-sized solids. Since the available pebble flux increases with the stellar metallicity and disk mass, \citet{Liu+2019} found that giant planet formation is more efficient around higher mass stars with higher metallicity. The solid core growth stops at the pebble isolation mass \citep{Lambrechts+2014b, Bitsch+2018}, where pebbles are stalled from drifting inward at the pressure bump formed by the protoplanet. Since the pebble isolation mass is smaller around lower mass stars, the formation of solid cores that are massive enough to trigger runaway gas accretion is less efficient around lower mass stars \citep{Liu+2019, Liu+2020}. To form giant planets around lower mass stars, the pebble accretion scenario requires multiple planet formation and giant impacts between them to grow further than the pebble isolation mass \citep{Pan+2024, Pan+2025}.


The pebble accretion models successfully reproduced the observed trend in the occurrence rate of giant exoplanets. However, these studies mainly focus on the formation of solid cores, and many simplifications are used for the gas accretion process. In the above pebble accretion models, protoplanets are assumed to start the Kelvin-Helmholtz contraction once the protoplanet reaches the pebble isolation mass. The Kelvin-Helmholtz contraction timescale depends on the boundary conditions of the planetary envelope \citep{Piso+2014} and the grain opacity \citep{Ikoma+2000}, which strongly affects the giant planet formation \citep{Bitsch+2021}. Previous studies have used the fixed grain opacity and neglected the effects of boundary conditions. Therefore, it is still unclear how these parameters affect the formation of cold Jupiters, especially in the pebble accretion scenario. 

\citet{Alibert+2011} pointed out that the formation rate of giant planets is related to the disk's lifetime since the protoplanets must undergo rapid gas accretion before disk dissipation. Despite the observations showing a large variety in the disk's lifetime \citep{Bayo+2012, Ribas+2015, Richert+2018}, how the occurrence rate of giant planets changes with the disk lifetime is not well discussed. While the detection of the inner dust ring in circumstellar disks shows that the disk's lifetime rarely depends on the stellar mass \citep{Richert+2018}, theoretical models for X-ray photoevaporation predict that the disk lifetime is longer around the lower mass stars because of the weaker X-ray irradiation \citep{Picogna+2021}. Additionally, long-lived accretion disks are detected around M-type stars, known as Peter Pan disks \citep{Silverberg+2020}. If the disk lifetime is longer around the lower mass stars, the giant planets may form in the core accretion scenario around M-dwarfs, even if the available solids are less abundant.

In this work, we model giant planet formation via pebble accretion while focusing on the crossover time, which is the time it takes for a protoplanet to reach the crossover mass.
We investigate how the crossover time changes with the formation environment and how the occurrence rate changes with stellar type. In Sec.~\ref{sec: method}, we describe our formation model. In Sec.~\ref{sec: result1}, we present the results of our numerical model and show how the crossover time changes with the formation environment. 
We discuss our results and model in Sec.~\ref{sec: discussion}. Our summary and conclusions are presented in Sec.~\ref{sec: conclusion}. 

\section{Methods}\label{sec: method}
\subsection{Disk model}

\begin{figure}
    \centering
    \includegraphics[width=1.\linewidth]{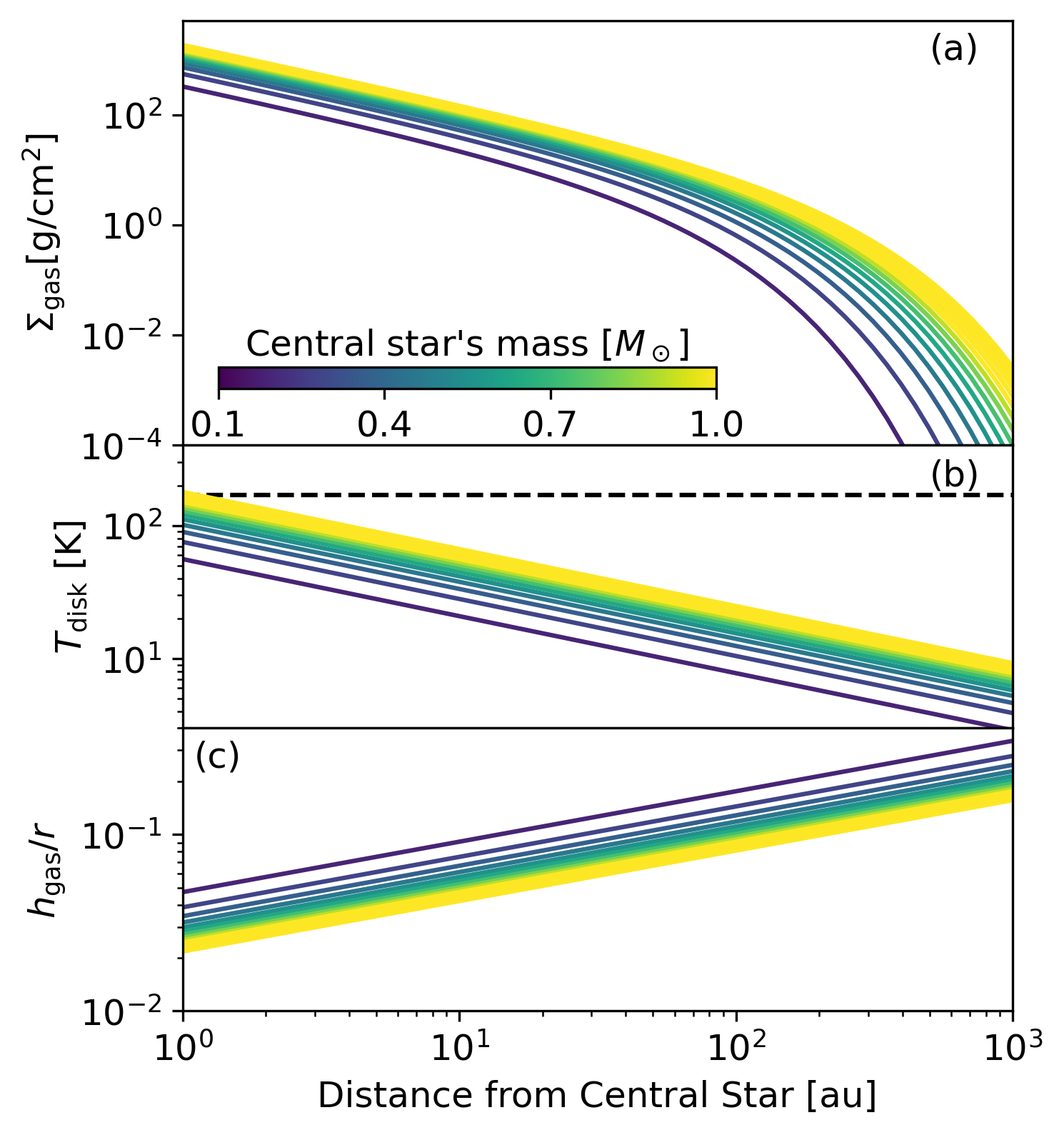}
    \caption{
    The disk model used in this work. Panel-(a), (b), and (c) show the surface density of the gaseous disk, disk temperature, and the disk's aspect ratio around different stellar masses as shown in the color bar in panel-(a). 
    The horizontal dashed line in panel-(b) shows $T_\mathrm{disk}= 170$ K, where H$_2$O condenses. 
    Here, we show the case when $t=10^5$ yr.
    }
    \label{fig: method_disk_model}
\end{figure}

Our disk model is based on the self-similar solution for the surface density profile of a gaseous disk \citep{Lynden-Bell+1974}.
We assume that the heating of the disk's mid-plane gas is dominated by irradiation from the central star rather than viscous heating, where cold Jupiters form. \revise{Assuming vertically optically thin and radially optically thick disk,} the mid-plane temperature $T_{\rm disk}$ is given by \revise{\citep[e.g.,][]{Chiang+1997, Ida+2016}}: 
\begin{align}
    T_{\rm disk} = 150 {\rm K} 
    \left( \frac{L_\mathrm{s}}{L_\odot} \right)^{2/7} 
    \left( \frac{M_\mathrm{s}}{M_\odot} \right)^{-1/7} 
    \left( \frac{r}{1 {\rm au}} \right)^{-3/7},
\end{align}
where $L_\mathrm{s}$ and $M_\mathrm{s}$ are the luminosity and mass of central star, and $r$ is the radial distance from the central star. If the disk viscosity scales with the radial distance as $\nu~\propto~r^\gamma$, the surface density of disk gas $\Sigma_\mathrm{gas}$ is given as: 
\begin{align}
    \Sigma_\mathrm{gas} = \frac{(2-\gamma) M_{\rm disk,0}}{2 \pi {R_{\rm disk}}^{2}} \left( \frac{r}{R_{\rm disk}} \right)^{-\gamma} T^{-\frac{\frac{5}{2}-\gamma}{2-\gamma}} \exp \left\{ -\frac{1}{T} \left(\frac{r}{R_{\rm disk}}\right)^{2-\gamma} \right\}, \label{eq:self-similar-solution}
\end{align}
with: 
\begin{align}
    \gamma &= \frac{3}{2} + \frac{{\rm d} \ln T_\mathrm{disk}}{{\rm d} \ln r} \\
    T &= 1 + \frac{t}{\tau_{\rm vis}}, \label{eq:normalized_time} \\
    \tau_{\rm vis} &= \frac{{R_{\rm d}}^2}{\nu_{\rm d}}, \label{eq:viscous_characteristic_timescale}
\end{align}
where $M_{\rm disk,0}$ is the disk total mass at $t=0$, $R_{\rm disk}$ is a radial scaling length of the disk, $\tau_{\rm vis}$ is the characteristic viscous timescale,  and $\nu_{\rm d}$ is a disk gas viscosity at $r=R_{\rm d}$. We use the $\alpha$-viscosity model \citep{Shakura+1973} and the disk gas viscosity is written as $\nu_\mathrm{d}=\alpha_{\rm acc} h^2_{\rm gas} \Omega_\mathrm{K}$, where $\alpha_{\rm acc}$ is the viscosity parameter, $h_{\rm gas}$ is the disk's scale height, and $\Omega_\mathrm{K}$ is the orbital angular velocity. 
\revise{
The disk's scale height is expressed as $h_{\rm gas}=c_\mathrm{s}/\Omega_\mathrm{K}$, where $c_\mathrm{s}$ is the sound speed of the disk gas. The sound speed is obtained using the ideal gas law, assuming a mean molecular weight of 2.34.
}
In this work, we use the fixed $\alpha_{\rm acc}=10^{-3}$. 

To investigate giant planet formation around different stellar masses, we introduce simple scaling laws following the numerical models available in the literature \citep{Miguel+2019, Burn+2021, Chachan+2023, Venturini+2024}.
Our scaling laws are given as:
\begin{align}
    L_\mathrm{s} &= a \left( \frac{M_\mathrm{s}}{M_\odot} \right)^{A}, \\
    M_\mathrm{disk,0} &= b \left( \frac{M_\mathrm{s}}{M_\odot} \right)^{B}, \\
    R_\mathrm{disk} &= c \left( \frac{M_\mathrm{disk, 0}}{0.1 M_\odot} \right)^{C}.
\end{align}
The coefficients we used in this work are summarized in the tab.~\ref{tab: scaling_laws}. Note that the coefficients are not strongly constrained by observations. For example, the mass-luminosity relation changes with the age of the stellar mass. Here, we begin with pre-main sequence stars and follow the numerical results obtained by \citet{Ramirez+2014, Choi+2016}. The relation between the disk's mass and stellar mass is taken from \citet{Andrews+2013}, and that between the disk's mass and size is taken from \citet{Andrews+2010}. %

\begin{table}
    \centering
    \begin{tabular}{ccc}
        \hline
        coefficient & value & refs. \\
        \hline
        $a$ & $1~L_\odot$ & - \\
        $b$ & $0.1~M_\odot$ & \citet{Andrews+2010}  \\
        $c$ & $100~{\rm au}$ & \citet{Andrews+2010} \\
        $A$ & 2 & \citet{Ramirez+2014} \\ 
        && \citet{Choi+2016} \\
        $B$ & 1 & \citet{Andrews+2013} \\
        $C$ & 1/1.6 & \citet{Andrews+2010} \\
        \hline
    \end{tabular}
    \caption{Coefficients of scaling laws.}
    \label{tab: scaling_laws}
\end{table}

\subsection{Pebble accretion model}
\revise{
Throughout this paper, we assume the protoplanet's eccentricity and inclination are negligibly small.
}
We adopt the pebble accretion model by \citet{Johansen+2017} where the pebble accretion rate is given by: 
\begin{align}
    \dot{M}_\mathrm{peb,acc} = \pi {R_\mathrm{peb,acc}}^2 \rho_\mathrm{p,mid} \bar{S} \delta v,
\end{align}
where $R_\mathrm{peb,acc}$ is the pebble accretion radius, $\rho_\mathrm{p,mid}$ is the pebbles' midplane density, $\Bar{S}$ is the stratification integral of pebbles, and $\delta v$ is the pebble's approaching speed which is given by $\Delta v +\Omega_\mathrm{K} R_\mathrm{p,acc}$ where $\Delta v$ is the sub-Keplerian speed. $R_\mathrm{peb,acc}$ is obtained by solving the equation: 
\begin{align}
    \tau_\mathrm{f} =\frac{\xi_\mathrm{B} \Delta v + \xi_\mathrm{H} \Omega_\mathrm{K} R_\mathrm{peb,acc}}{\mathcal{G} M_\mathrm{p}/{R_\mathrm{peb,acc}}^2},
\end{align}
where $\tau_\mathrm{f}$ is the Stokes number, $\mathcal{G}$ is the gravitational constant, and $\xi_\mathrm{B}$ and $\xi_\mathrm{H}$ are fitting parameters. Numerical results are well reproduced with $\xi_\mathrm{B}=\xi_\mathrm{H}=0.25$ \citep{Ormel+2010, Lambrechts+2012}. The stratification integral $\Bar{S}$ is given by: 
\begin{align}
    \bar{S} = \frac{1}{\pi{R_\mathrm{peb,acc}^2}} \int_{z_\mathrm{p}-R_\mathrm{peb,acc}}^{z_\mathrm{p}+R_\mathrm{peb,acc}} 2 \exp \left( -\frac{z^2}{2 {h_\mathrm{peb}}^2} \right) \sqrt{{R_\mathrm{peb,acc}}^2-\left( z-z_\mathrm{p} \right)^2} \mathrm{d} z,
\end{align}
where $z_\mathrm{p}$ is the height of the protoplanet measured from the disk's mid-plane, and $h_\mathrm{peb}$ is the scale height of the pebble layer. We set $z_\mathrm{p}=0$ in this study.

The mid-plane density of pebbles is $\rho_\mathrm{p, mid}=\Sigma_\mathrm{peb}/\sqrt{2} h_\mathrm{peb}$ where $\Sigma_\mathrm{peb}$ is the surface density of pebbles, which is given by: 
\begin{align}
    \Sigma_\mathrm{peb} &= \frac{ \dot{M}_\mathrm{peb}}{2 \pi r v_\mathrm{peb}}, 
\end{align}
\revise{
with the radial velocity of pebbles:
\begin{align}
    v_\mathrm{peb} &= \frac{2 \eta v_\mathrm{K}}{\tau_\mathrm{f}+{\tau_\mathrm{f}}^{-1}}+\frac{\nu}{r},
\end{align}
and:
\begin{align}
    \eta &= - \frac{1}{2} \left( \frac{h_\mathrm{gas}}{r} \right)^2 \frac{\partial \ln P_\mathrm{disk}}{\partial \ln r}.
\end{align}
Here,
$\dot{M}_\mathrm{peb}$ is the pebble flux, $v_\mathrm{K}$ is the Kepler velocity at the protoplanet's orbit, and $P_\mathrm{disk}$ is the gas pressure of the disk, which is calculated assuming an ideal gas.
}
The \revise{pebble layer's} scale height $h_\mathrm{peb}$ is: 
\begin{align}
    h_\mathrm{peb} = h_\mathrm{gas} \sqrt{\frac{\alpha_\mathrm{turb}}{\tau_\mathrm{f}}}, \label{eq: Hpeb}
\end{align}
where $\alpha_\mathrm{turb}$ is the local turbulent viscosity. 
\revise{
Note that this equation is valid in the regime $\alpha_\mathrm{turb}\ll\tau_\mathrm{f}$.
}
The local turbulent viscosity $\alpha_\mathrm{turb}$ could be different from the accretion viscosity $\alpha_\mathrm{acc}$, which drives the global angular momentum transfer \citep[e.g.][]{Bai+2016}. Following \citet{Ida+2018}, we assume $\alpha_\mathrm{acc} \geq \alpha_\mathrm{turb}$ and set $\alpha_\mathrm{turb}$ as an input parameter. The pebble's size is obtained by equating the pebble growth timescale with the drift timescale \citep{Lambrechts+2014}. The surface density of pebbles transforms into: 
\begin{align}
    \Sigma_\mathrm{peb} &= \sqrt{\frac{2 \dot{M}_\mathrm{peb} \Sigma_\mathrm{gas}}{\sqrt{3} \pi \epsilon_\mathrm{p} r v_\mathrm{K}} },
\end{align}
where $\epsilon_\mathrm{p}$ is the pebble's sticking efficiency which we set to $0.5$. The pebble flux $\dot{M}_\mathrm{peb}$ is given by the model developed by \citet{Lambrechts+2014}: 
\begin{align}
    \dot{M}_\mathrm{peb} = 2 \pi r_\mathrm{g} \frac{\mathrm{d} r_\mathrm{g}}{\mathrm{d} t} Z_\mathrm{disk} \Sigma_\mathrm{gas} (r=r_\mathrm{g},t=0) \label{eq: peb_flux}
\end{align}
with
\begin{align}
    r_\mathrm{g} &= \left( \frac{3}{16} \right)^{1/3}  \left( \mathcal{G} M_\mathrm{s} \right)^{1/3} \left(\epsilon_\mathrm{D} Z_\mathrm{disk} \right)^{2/3} t^{2/3}, \label{eq: pebble_production_line}\\
    \frac{\mathrm{d} r_\mathrm{g}}{\mathrm{d} t} &= \frac{2}{3} \left( \frac{3}{16} \right)^{1/3}  \left( \mathcal{G} M_\mathrm{s} \right)^{1/3} \left(\epsilon_\mathrm{D} Z_\mathrm{disk} \right)^{2/3} t^{-1/3},
\end{align}
\revise{
where $Z_\mathrm{disk}$ is the disk's metallicity, which is set to 0.01 throughout this paper, and $\epsilon_\mathrm{D}=0.05$.
}
\revise{
Note that in Eq.~\ref{eq: peb_flux} we use the surface density of solids at $t=0$ and neglect the radial movement of small dust with the viscously evolving gas. We adopt this model for the mass conservation of solids because our model does not include dust crossing of the pebble condensation front by advection and diffusion. Therefore, our model overestimates the pebble flux if we calculate it using the time-evolving $\Sigma_\mathrm{gas}$. More details on that point can be found in Appendix~\ref{app: disk_model}.
}

Pebble accretion terminates when the growing planet reaches  the pebble isolation mass \citep{Lambrechts+2014b, Bitsch+2018}, which is given by: 
\begin{eqnarray}    
    M_\mathrm{iso} &=& 25.0 M_\oplus 
        \left( \frac{M_\mathrm{s}}{M_\odot} \right) 
        \left( \frac{h/r}{0.05}\right)^3 \nonumber \\
        && \times \left( 0.34 \left[ \frac{-3}{\log_{10}(\alpha_\mathrm{turb})}\right]^4+ 0.66\right) 
        \left(1 - \frac{ \frac{\partial \ln P_\mathrm{disk}}{\partial \ln r}+2.5}{6}  \right). \label{eq: Miso_peb_Bitsch}
\end{eqnarray}
We introduce an exponential cutoff at the pebble isolation mass and a cap for the pebble flux, which limits the heavy-element accretion rate, and the solid accretion rate is finally given by: 
\begin{eqnarray}    
    \dot{M}_\mathrm{Z} = \min \left( \dot{M}_\mathrm{peb}, \dot{M}_\mathrm{peb,acc} \exp{ \left(- \left[ \frac{M_\mathrm{p}}{M_\mathrm{iso}} \right]^{10} \right)} \right),
\end{eqnarray}
where $M_\mathrm{p}$ is the protoplanet's mass.

\subsection{Gas accretion}
For calculating the gas accretion rate onto the growing protoplanet, we use the MESA-extention code developed by \citet{Valletta+2020}, which is based on the stellar evolution code MESA \citep{Paxton+2011, Paxton+2018}. MESA's version is r24.03.1. The planet is assumed to be in hydrostatic equilibrium and spherically symmetric. The equations of state are adapted from \citet{Muller+2020}. We assume that the outer edge of the planetary envelope is connected to the local disk gas. The surface temperature $T_\mathrm{surf}$ and pressure $P_\mathrm{surf}$ are set to $T_\mathrm{disk}$ and $P_\mathrm{disk}$. 
\revise{
Before entering the runaway gas accretion ($M_\mathrm{p}\lesssim10M_\oplus$), the protoplanet's radius is comparable to or smaller than the disk's scale height (see Eq.~\ref{eq: gas_acc_radius} below). In this regime, temperature and pressure gradients across the surface of the envelope are relatively small, justifying the assumption of spherical symmetry within the scope of this study.
}

The gas accretion rate $\dot{M}_\mathrm{env}$ is calculated by the method developed in \citet{Pollack+1996}. In each timestep, we add a mass $\Delta m$ to fill the gap between the planetary radius $R_\mathrm{p}$ and the gas accretion radius $R_\mathrm{gas,acc}$, which is given by: 
\begin{equation}
    R_\mathrm{gas,acc} = \frac{\mathcal{G} M_\mathrm{p}}{ {c_\mathrm{s}}^2 + 4 \mathcal{G} M_\mathrm{p}/R_\mathrm{H}}, \label{eq: gas_acc_radius}
\end{equation}
where $R_\mathrm{H}$ is the Hill radius. 
By adding $\Delta m$, $R_\mathrm{p}$ increases (or decreases if $\Delta m$ is negative), and the gap becomes narrower. We iterate each step until $\delta = \left| (R_\mathrm{p}-R_\mathrm{gas,acc}) / R_\mathrm{gas,acc} \right|$ becomes smaller than $10^{-3}$.


We use a dust grain opacity prescription of \citet{Valencia+2013} where the grain opacity is scaled with a grain opacity factor $f_\mathrm{g}$. $f_\mathrm{g}=1$ corresponds to the interstellar medium's grain opacity, while in planetary atmospheres, lower values are expected \citep{Movshovitz+2010, Mordasini2014, Ormel2014}. In this study, we set $f_\mathrm{g}$ as a free parameter.

It is known that the heavy-element deposition in the planetary envelope accelerates the gas accretion process \citep{Hori+2010, Hori+2011, Venturini+2015, Venturini+2016, Valletta+2020, Mol-Lous+2024}. For simplicity, in this study we assume that the accreted heavy elements settle to the core and release the accretion energy there.
Instead, we focus on the disk conditions and how the stellar mass affects the formation of giant planets. We discuss the importance of heavy-element deposition on giant planet formation in sec.~\ref{sec: discussion_deposition}.

Convection is modeled using the mixing-length theory. The mixing length parameter $\alpha_\mathrm{MLT}$ is set to 2, which is estimated from the stellar evolution models \citep{Paxton+2011}. The value $\alpha_\mathrm{MLT}$ in planetary condition is poorly constrained, and $\alpha_\mathrm{MLT}$ could be significantely lower \citep{Leconte+2012}. However, gas accretion is rather independent of $\alpha_\mathrm{MLT}$ if the envelope is not enriched with heavy elements (see Appendix~\ref{sec: discussion_mixing_length}). Therefore, in our baseline simulations, we use $\alpha_\mathrm{MLT}=2$.

\subsection{Formation simulation setup}
We start the simulation assuming a heavy-element core with a mass $M_0=0.01 M_\oplus$ at $t=t_0$. The location of the protoplanet $r$ is fixed at the same orbital period, which corresponds to 3 au around a solar mass star; namely, $r=3 \mathrm{au}~{\left(M_\mathrm{s}/M_\odot\right)}^{1/3}$. 
We neglect the planetary migration in our simulations. To form cold Jupiters, protoplanets need to enter the outward migration region \citep{Coleman+2016}, or start their formation in the orbit farther than $10$ au \citep{Bitsch+2015}. Otherwise, rapid type-I migration allows the protoplanets to migrate to the disk's inner edge and become warm/hot-Jupiters. The migration speed and the location of the outward migration region depend on various parameters such as the disk's structure \citep{Tanaka+2002, Coleman+2016}, temperature profile, and the corotation torque \citep{Paardekooper+2010, Paardekooper+2011, Paardekooper+2014}. We fix the protoplanet's orbit to reduce the number of free parameters and focus on the gas accretion process.
\revise{
The effects of planetary migration are discussed in sec.~\ref{sec: discussion_migration}. 
}
The heavy-element core increases its mass by accreting pebbles and disk gas. The simulation is terminated when the protoplanet reaches the crossover mass, i.e., when the planetary envelope mass $M_\mathrm{env}$ equals the core mass $M_\mathrm{core}$, or when the protoplanet's growth timescale $\tau_\mathrm{grow} = M_\mathrm{p}/\dot{M}_\mathrm{p}$ exceeds $10^9$ yr.

We perform parameter studies when changing the stellar mass $M_\mathrm{s}$, core formation time $t_\mathrm{0}$, grain opacity factor $f_\mathrm{g}$, and the disk's turbulent viscosity $\alpha_\mathrm{turb}$. 
\revise{
We summarise the parameter ranges used in this paper in Tab.~\ref{tab: parameters}.
}

\begin{table}
    \centering
    \begin{tabular}{cccc}
        \hline
        parameter & & range & result \\
        \hline
        grain opacity factor & $f_\mathrm{g}$ & $10^{-3}$-$10^{0}$ & Sec.~\ref{sec: result_psfg} \\
        turbulent viscosity & $\alpha_\mathrm{turb}$ & $10^{-5}$-$10^{-3}$ & Sec.~\ref{sec: result_psAlpha} \\
        core formation time & $t_0$ & $10^5$-$10^7$ yr & Sec.~\ref{sec: result_psTs} \\
        central star's mass & $M_\mathrm{s}$ & $0.1$-$1.5$ $M_\odot$ & Sec.~\ref{sec: result_psTs} \\
        \hline
    \end{tabular}
    \caption{
    \revise{
    Parameters used in this study.
    }
    }
    \label{tab: parameters}
\end{table}

\section{Results}\label{sec: result1}

\subsection{The role of grain opacity}\label{sec: result_psfg}
\begin{figure}
    \centering
    \includegraphics[width=0.8\linewidth]{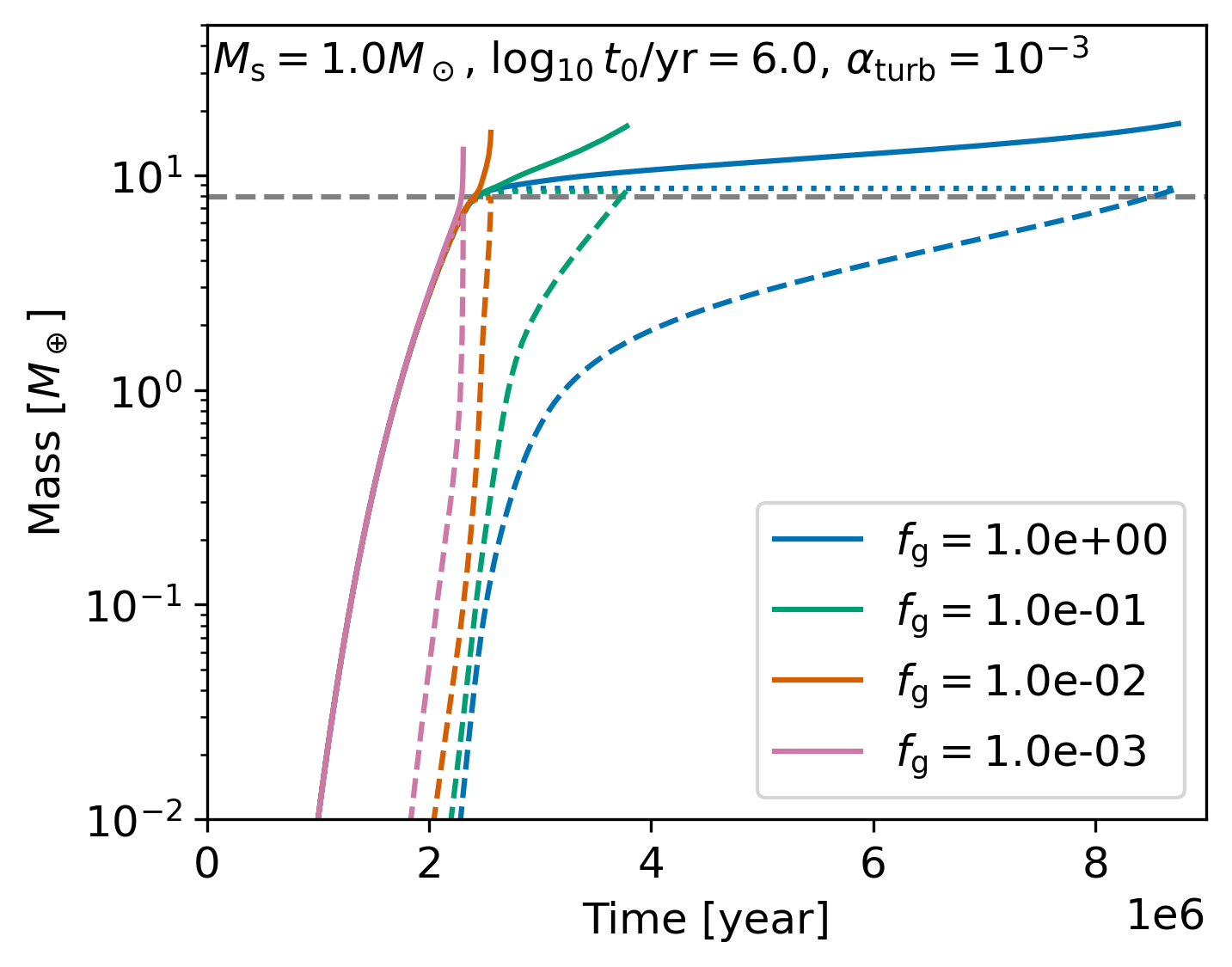}
    \caption{
    Growth of protoplanets up to crossover mass around $1 M_\odot$ star with the core formation time $t_0=1$ Myrs. Different colors show the cases with different grain opacity $f_\mathrm{g}$. The solid, dashed, and dotted lines show the total, envelope, and solid core mass, respectively. The horizontal gray dashed line shows the pebble isolation mass. 
    }
    \label{fig: result_psfg}
\end{figure}

Figure~\ref{fig: result_psfg} shows the planetary formation when assuming different grain opacity factors $f_\mathrm{g}$. 
Before reaching the pebble isolation mass (horizontal dashed line), the protoplanet's mass increases mainly via pebble accretion ({\it core growth phase}). Since the rapid pebble accretion brings energy large enough to support the hydrostatic condition, the protoplanet's envelope doesn't contract. The gas accretion rate is much lower than the solid accretion rate and is controlled by the expansion of the accretion radius $R_\mathrm{gas,acc}$. 
Once the protoplanet reaches pebble isolation mass, the pebble accretion stops, and the envelope begins to contract. The cooling timescale controls the gas accretion rate ({\it contraction phase}). Since the cooling timescale depends on the atmospheric opacity, the gas accretion rate is shorter for the lower grain opacity factors. 

\begin{figure}
    \centering
    \includegraphics[width=0.8\linewidth]{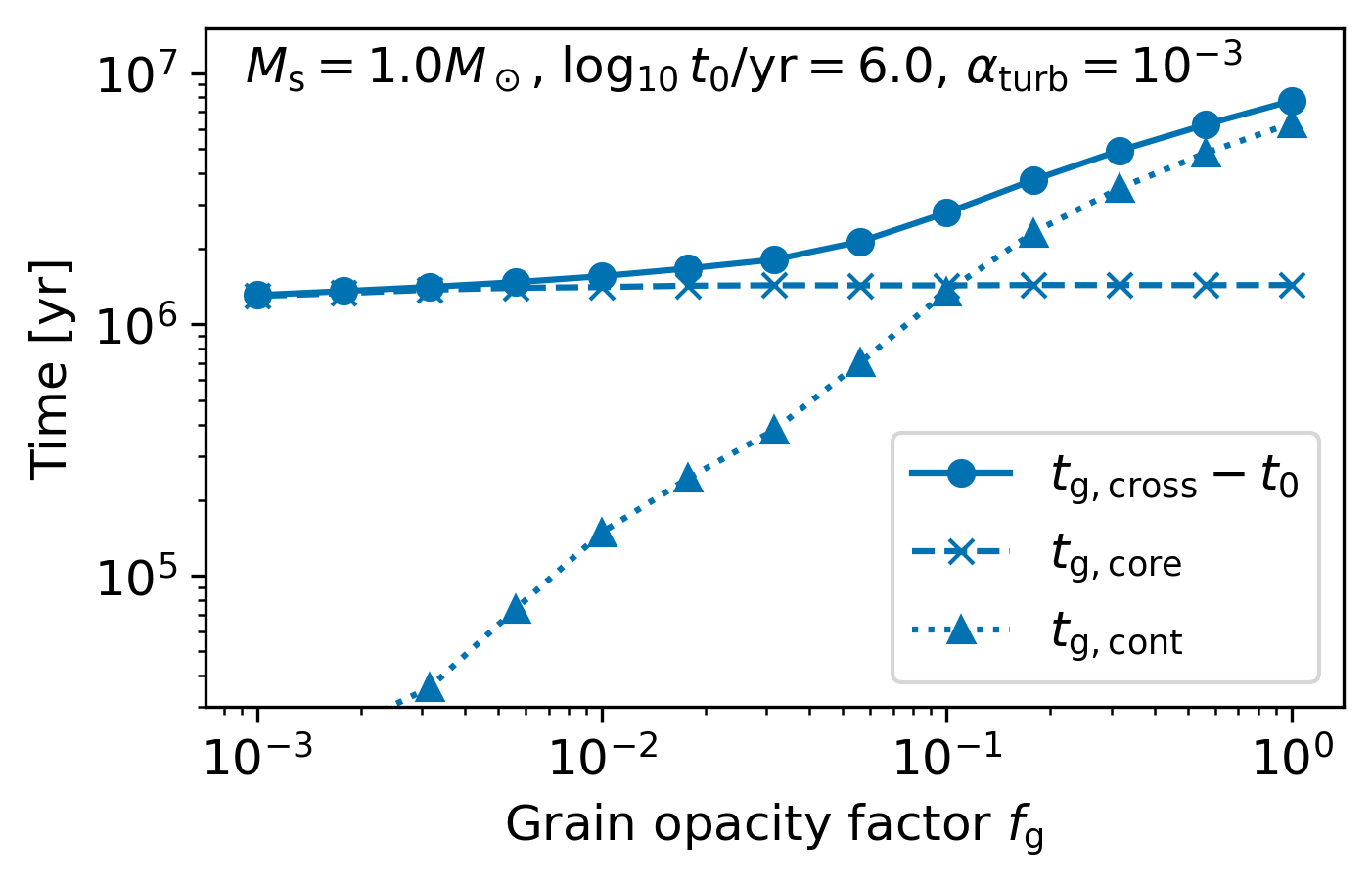}
    \caption{
    Growth times obtained in the simulations with $M_\mathrm{s}=1M_\odot$, $t_0=10^6$ yr, and $\alpha_\mathrm{turb}=10^{-3}$. 
    The solid line shows the crossover time, \revise{measured from the core formation time}. The dashed and dotted lines show the times required for the protoplanet to grow from $M_\mathrm{0}$ to $M_\mathrm{iso}$ and from $M_\mathrm{iso}$ to $M_\mathrm{cross}$, respectively.
    }
    \label{fig: time_growth_psfg}
\end{figure}

Figure~\ref{fig: time_growth_psfg} shows 
\revise{
the crossover time $t_\mathrm{g, cross}$, defined as the time when the protoplanet reaches the crossover mass $M_\mathrm{cross}$ (solid lines).
For comparison, we also plot the core growth time $t_\mathrm{g, core}$, defined as the time required for the protoplanet to grow from the initial mass $M_0$ to the pebble isolation mass $M_\mathrm{iso}$ (dashed lines). Additionally, the envelope's contraction time $t_\mathrm{g, cont}$, representing the time between reaching $M_\mathrm{iso}$ and $M_\mathrm{cross}$, is shown with dotted lines. Note that in Fig.~\ref{fig: time_growth_psfg} —as well as Fig.~\ref{fig: time_growth_psAlp} and \ref{fig: time_growth_pstCore}— we plot the crossover time measured from the core formation time $t_\mathrm{g, cross}-t_0$ to facilitate direct comparison with the core growth time $t_\mathrm{g, core}$ and the contraction time $t_\mathrm{g, cont}$.
}
Here, we show the cases with $M_\mathrm{s}=1M_\odot$, $t_0=10^6$ yr, and $\alpha_\mathrm{turb}=10^{-3}$. The crossover time increases with $f_\mathrm{g}$ because a larger $f_\mathrm{g}$ delays the contraction of the planetary envelope. Unlike in the contraction phase, the core growth phase is independent of $f_\mathrm{g}$.
If $f_\mathrm{g}$ is greater than 0.1, we find that the contraction phase becomes longer than the core growth phase; therefore, the crossover time depends on $f_\mathrm{g}$. If $f_\mathrm{g}$ is smaller than 0.1, the protoplanet reaches the crossover mass shortly after the pebble isolation mass is reached. In this case, the formation of giant planets is regulated only by pebble accretion, and the crossover time rarely depends on $f_\mathrm{g}$.

\subsection{The role of turbulent viscosity}\label{sec: result_psAlpha}

\begin{figure}
    \centering
    \includegraphics[width=0.8\linewidth]{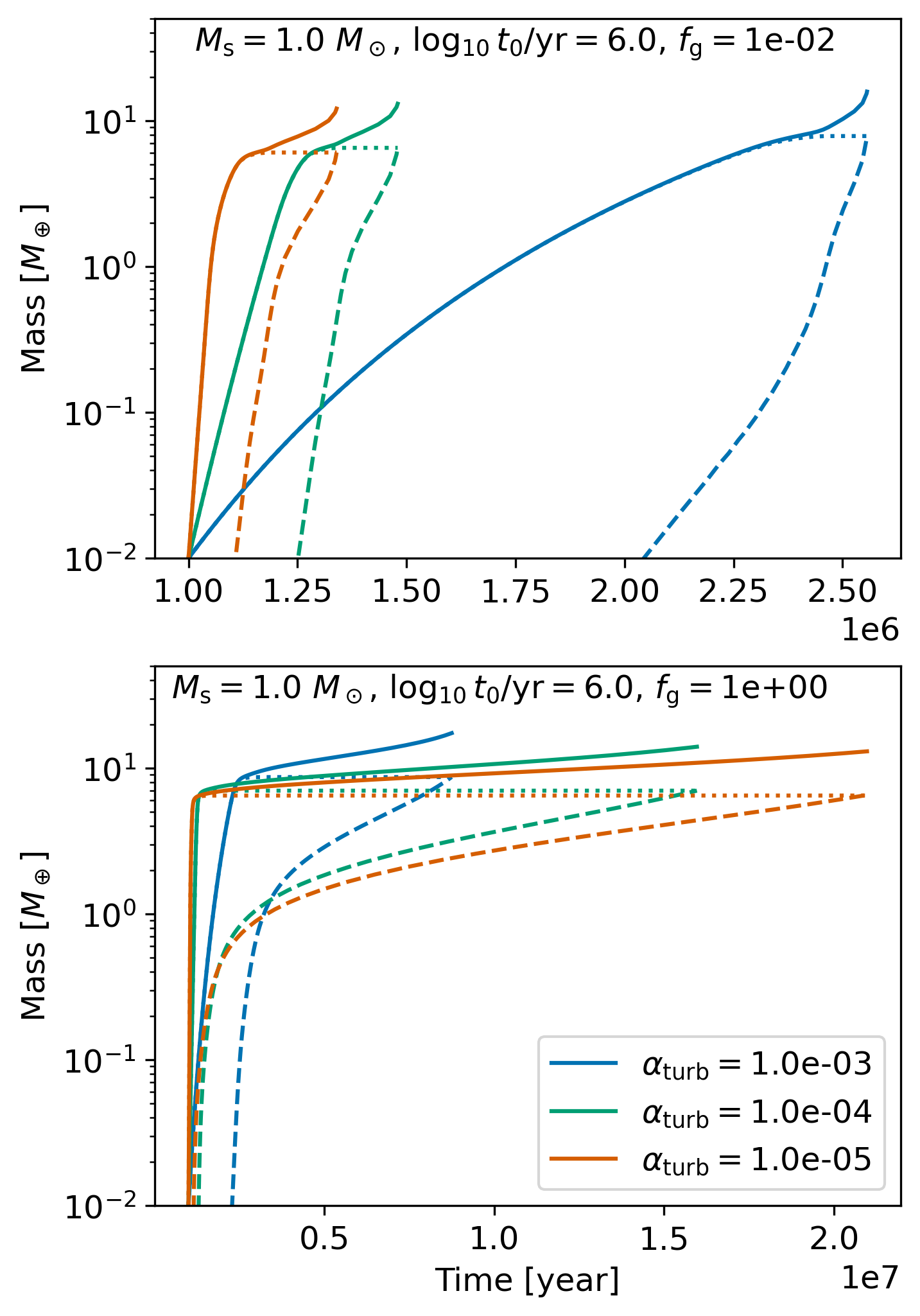}
    \caption{
    Same as Fig.~\ref{fig: result_psfg}, but for different assumed turbulent viscosities $\alpha_\mathrm{turb}$. The upper and lower panels show the cases with $f_\mathrm{g}=10^{-2}$ and $f_\mathrm{g}=10^{0}$, respectively.
    }
    \label{fig: result_psAlp}
\end{figure}

\begin{figure}
    \centering
    \includegraphics[width=0.8\linewidth]{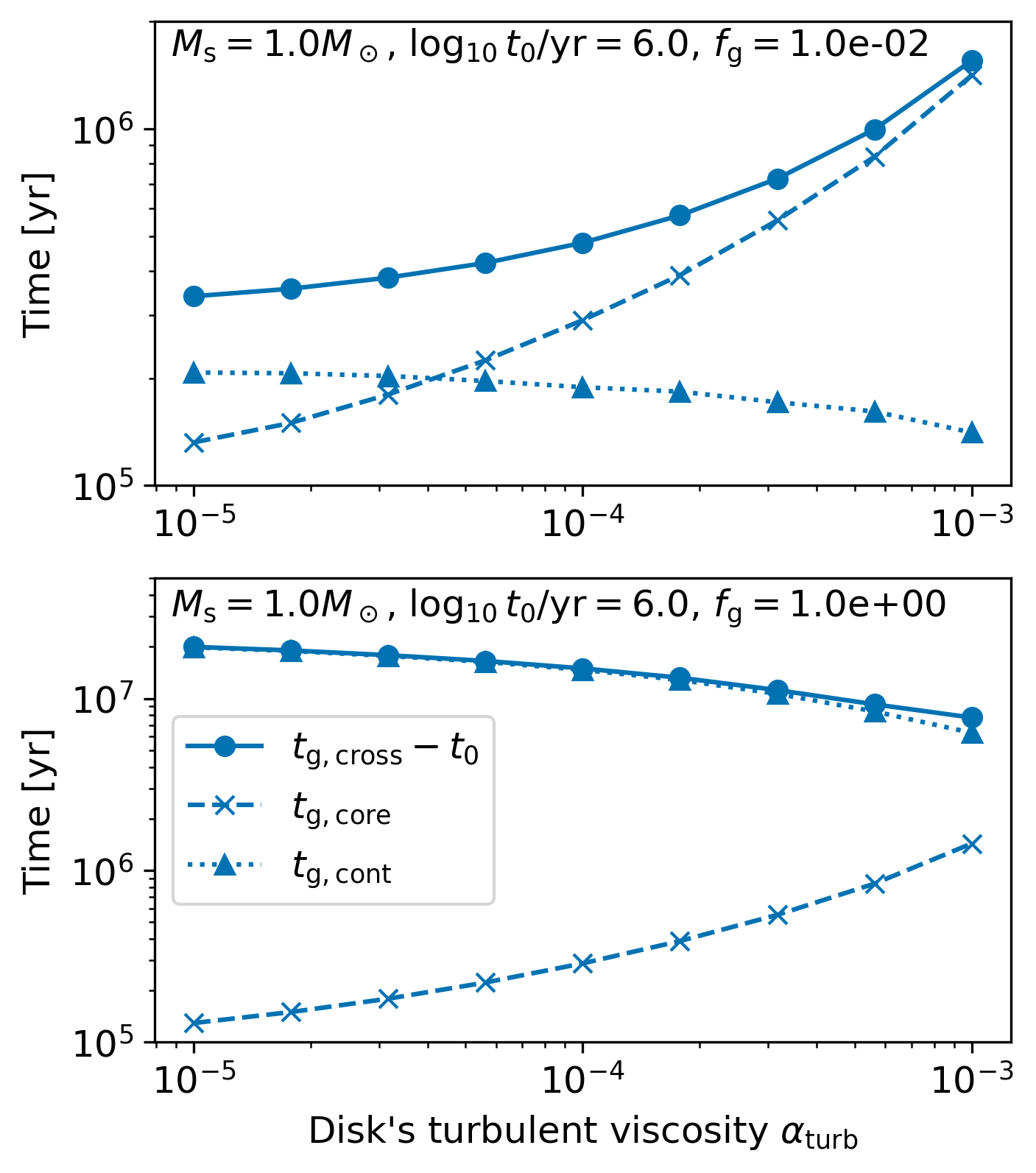}
    \caption{
    Same as Fig.~\ref{fig: time_growth_psfg}, but for different assumed turbulent viscosities $\alpha_\mathrm{turb}$. The upper and lower panels show the cases with $f_\mathrm{g}=10^{-2}$ and $f_\mathrm{g}=10^{0}$, respectively.
    }
    \label{fig: time_growth_psAlp}
\end{figure}

Next, we explore the importance of the assumed disk's turbulent viscosity. The disk's turbulent viscosity affects the pebble accretion rate and isolation mass. The pebble accretion rate is higher for lower $\alpha_\mathrm{turb}$ because of the smaller \revise{pebble layer's} scale height (see Eq.~\ref{eq: Hpeb}). On the other hand, the pebble isolation mass becomes smaller for lower $\alpha_\mathrm{turb}$ as shown in Eq.~\ref{eq: Miso_peb_Bitsch}. Figure~\ref{fig: result_psAlp} shows the time evolution of protoplanets with different $\alpha_\mathrm{turb}$. We also plot the crossover time presented in Fig.~\ref{fig: time_growth_psAlp}. Here, we show the cases with $M_\mathrm{s}=1M_\odot$, $t_0=10^6$ yr, and $f_\mathrm{g}=0.01$ (upper panels) and $1$ (lower panels). Due to the higher pebble accretion rate, the protoplanet reaches the pebble isolation mass earlier for lower $\alpha_\mathrm{turb}$. The planetary core mass is smaller for lower $\alpha_\mathrm{turb}$ because of the smaller pebble isolation mass. As shown in Fig.~\ref{fig: time_growth_psAlp}, the core growth time $t_\mathrm{g,core}$ is shorter for lower $\alpha_\mathrm{turb}$. On the other hand, the contraction time $t_\mathrm{g, cont}$ is longer for lower $\alpha_\mathrm{turb}$. This is because of the smaller pebble isolation mass. The envelope's contraction timescale strongly depends on the mass of protoplanets \citep{Ikoma+2000, Piso+2014}.

When the grain opacity is small (e.g., $f_\mathrm{g}=0.01$), the core growth time regulates the crossover time as discussed in Sec.~\ref{sec: result_psfg}, and the crossover time decreases for lower $\alpha_\mathrm{turb}$. On the other hand, when $f_\mathrm{g}=1$, the dominant growth phase is the envelope's contraction phase, and the crossover time is longer for lower $\alpha_\mathrm{turb}$.



\subsection{Core formation time and stellar mass}\label{sec: result_psTs}



\begin{figure}
    \centering
    \includegraphics[width=0.8\linewidth]{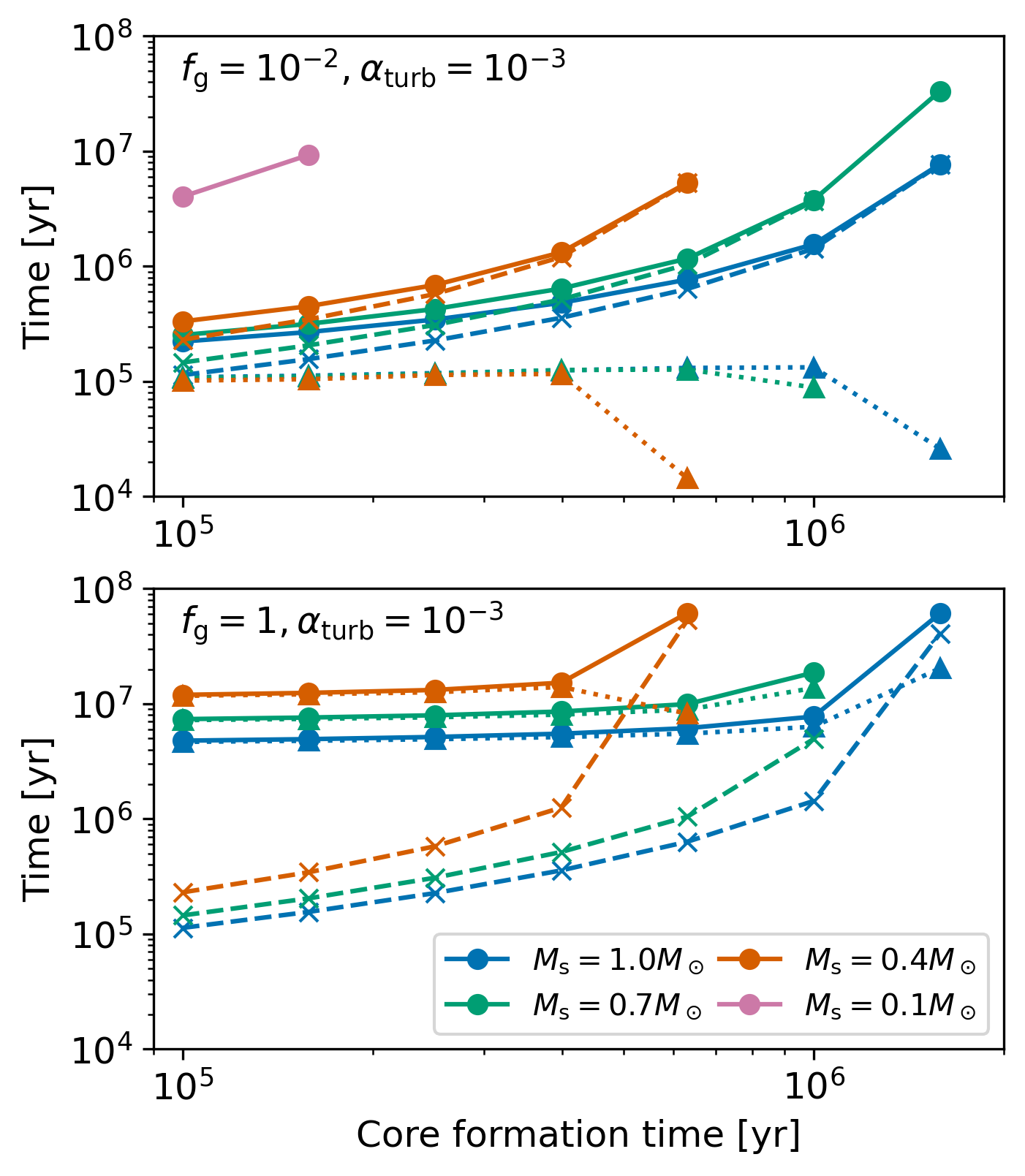}
    \caption{
    Same as Fig.~\ref{fig: time_growth_psfg}, but for different core formation times $t_0$. \revise{The top and bottom} panels show the cases with $f_\mathrm{g}=0.01$ and $0.1$, respectively. The different colors correspond to different stellar masses. The solid, dashed, and dotted lines show the crossover time \revise{measured from the core formation time $t_\mathrm{g, cross}-t_0$}, the core growth time $t_\mathrm{g, core}$, and the contraction time $t_\mathrm{g, cont}$, respectively. Here, we show the cases with $\alpha_\mathrm{turb}=10^{-3}$.
    }
    \label{fig: time_growth_pstCore}
\end{figure}

Figure~\ref{fig: time_growth_pstCore} shows the crossover time as a function of the core formation time. We show the cases with low grain opacity $f_\mathrm{g}=0.01$ and high grain opacity $f_\mathrm{g}=1$ in \revise{the top and bottom panels}, respectively. First, we focus on the cases around $1 M_\odot$ stars (blue lines). We find that the protoplanets take longer to reach the pebble isolation mass (dashed lines) if the core formation time is longer. This is because both the total dust mass remaining in the protoplanetary disk and the pebble flux decrease with time. Unlike in the core growth phase, the growth time in the contraction phase $t_\mathrm{g, cont}$ (dotted lines) depends only weakly on $t_0$. 

We also plot the cases around different stellar masses in Fig.~\ref{fig: time_growth_pstCore}. Because of the lower pebble flux, the protoplanet takes longer to reach the pebble isolation mass around lower-mass stars. When $f_\mathrm{g}=0.01$, the protoplanet reaches crossover mass soon after the pebble isolation mass is reached, and $t_\mathrm{g, cont}$ is rather independent of the mass of the host stars. However, $t_\mathrm{g,cont}$ depends on the stellar type when $f_\mathrm{g}=1$. The pebble isolation mass is larger for planets around more massive stars. Due to the larger core mass, the planetary envelope contracts quickly, and therefore, $t_\mathrm{g, cont}$ is shorter around more massive stars. 

We find that protoplanets do not reach pebble isolation around very low mass stars $M_\mathrm{s}=0.1M_\mathrm{s}$. However, the protoplanets could reach the crossover mass if the core formation time is as short as $10^5$ yrs and $f_\mathrm{g}=0.01$. In the case of $t_0=10^5$ yr, the protoplanet's core mass is $1.2 M_\oplus$. Despite the small core mass, the protoplanet can accrete disk gas due to the low atmospheric opacity. With higher grain opacity $f_\mathrm{g}=1$, protoplanets do not reach the crossover mass with $M_\mathrm{s}=0.1M_\mathrm{s}$.

While the crossover time $t_\mathrm{g, cross}$ ranges from $10^5$ to $10^7$ yrs for lower grain opacity cases $f_\mathrm{g}=0.01$, $t_\mathrm{g, cross}$ exceeds $5$ Myrs for the higher grain opacity cases $f_\mathrm{g}=1.0$. Since observations suggest that typical disk lifetimes are $\sim1-3$ Myrs \citep{Richert+2018}, the frequency of giant planet formation is expected to be low if $f_\mathrm{g}=1.0$. In the next section, we further discuss the occurrence rate of giant planets.

\subsection{Crossover time}\label{sec: result_crossover}

\begin{figure}
    \centering
    \includegraphics[width=1.\linewidth]{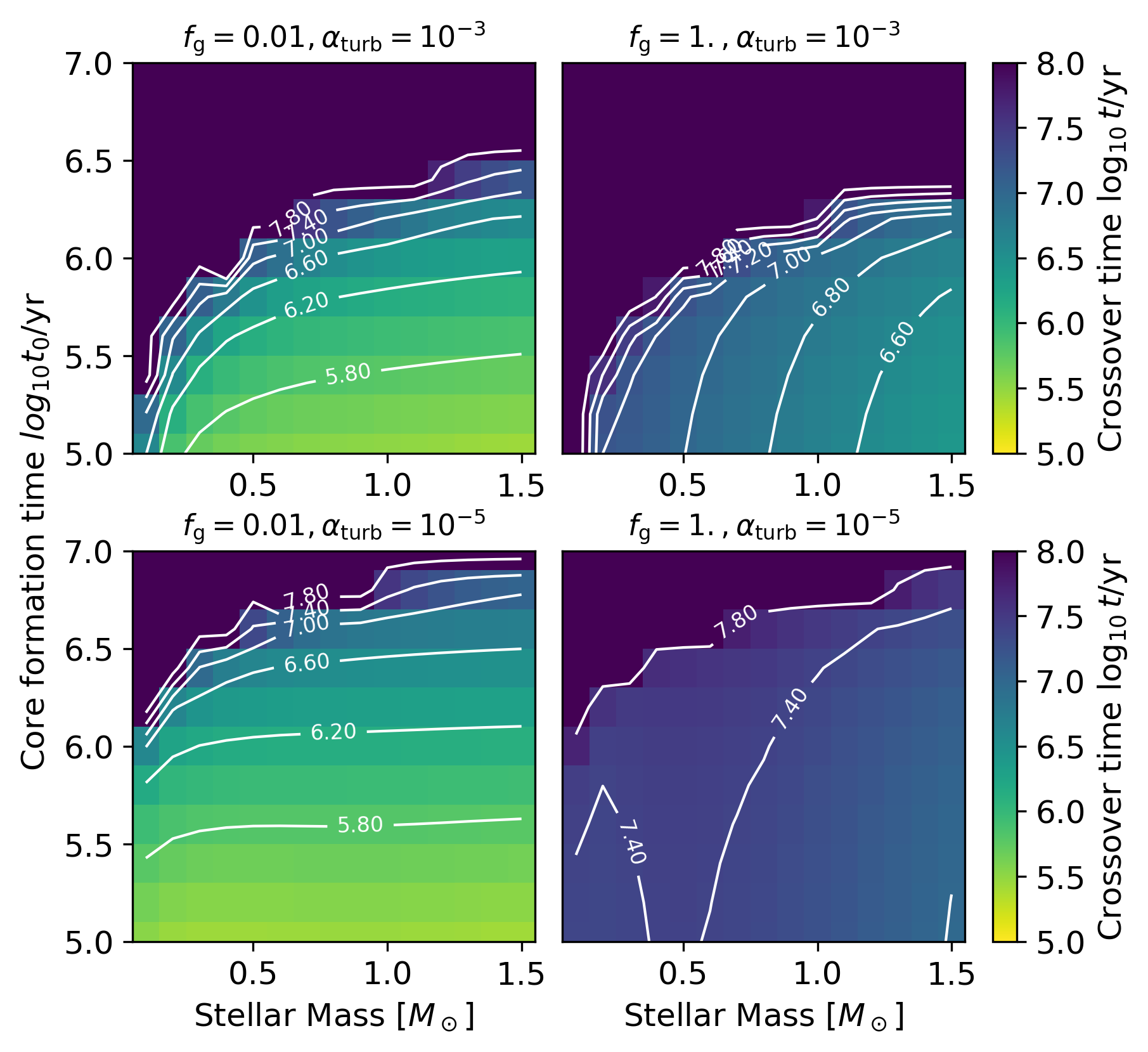}
    \caption{
    The crossover time obtained from simulations starting with an initial core of $M_\mathrm{core}=0.01 M_\oplus$. We plot the obtained crossover time $t_\mathrm{g,cross}$ as a color contour, and the color bar is shown on the right side of the panels. The horizontal and vertical axes are stellar mass $M_\mathrm{s}$ and the core formation time $t_\mathrm{0}$. The black lines show the contour lines and the number on each line shows $\log_\mathrm{10} t_\mathrm{g,cross}$. The left and right columns correspond to the low ($f_\mathrm{g}=0.01$) and high ($f_\mathrm{g}=1$) grain opacity cases. The top and bottom rows show the high ($\alpha_\mathrm{turb}=10^{-3}$) and low ($\alpha_\mathrm{turb}=10^{-5}$) turbulent viscosity cases. 
    }
    \label{fig: result_Mc001}
\end{figure}

Figure~\ref{fig: result_Mc001} shows the inferred crossover time $t_\mathrm{g, cross}$ as a function of stellar mass $M_\mathrm{s}$ and core formation time $t_0$, for different $f_\mathrm{g}$ and $\alpha_\mathrm{turb}$. 

The crossover time shortens with faster core formation around more massive stars. When $f_\mathrm{g}=1.0$ and $\alpha_\mathrm{turb}=10^{-3}$, giant planet formation is rather slow, and crossover time exceeds 3 Myrs. 
In that case, only the long-lived protoplanetary disks would enable the formation of giant planets. 
When the grain opacity factor is as small as $f_\mathrm{g}=0.01$, the crossover time becomes $\sim1$ Myr. Giant planets would form if the initial core formation ends within 1 Myrs, except around late M-dwarfs ($M_\mathrm{s}=0.1 M_\odot$). 

As shown in Sec.~\ref {sec: result_psAlpha}, a smaller turbulent viscosity of $\alpha_\mathrm{turb}=10^{-5}$ shortens the crossover time if the grain opacity is small as $f_\mathrm{g}=0.01$, but lengthens if $f_\mathrm{g}=1$. A smaller turbulent viscosity $\alpha_\mathrm{turb}=10^{-5}$ and grain opacity $f_\mathrm{g}=0.01$ are required to form giant planets around late M-dwarfs. In this case, the crossover time weakly depends on the stellar type if the initial core forms within 1 Myr. Even if the initial core formation takes $t_0\sim3$ Myrs ($\log_\mathrm{10} t_\mathrm{g, cross}=6.5$), giant planets could still form around high-mass stars with long-lived protoplanetary disks. On the other hand, no giant planets would form with a low turbulent viscosity $\alpha_\mathrm{turb}=10^{-5}$ and a high grain opacity $f_\mathrm{g}=1$.

\subsection{From core formation to crossover}\label{sec: result2}

\revise{
In the above sections,
}
we investigate the formation of giant planets, setting the core formation time $t_0$ as an input parameter. 
\revise{
Hereafter,
}
we estimate $t_0$ by calculating the pebble accretion onto a protoplanet, which is set to be the largest planetesimal formed by streaming instability. Then, we estimate the crossover timescale using the numerical results obtained in Sec.~\ref{sec: result_crossover}.

\subsubsection{Core formation time}

\begin{figure}
    \centering
    \includegraphics[width=1.\linewidth]{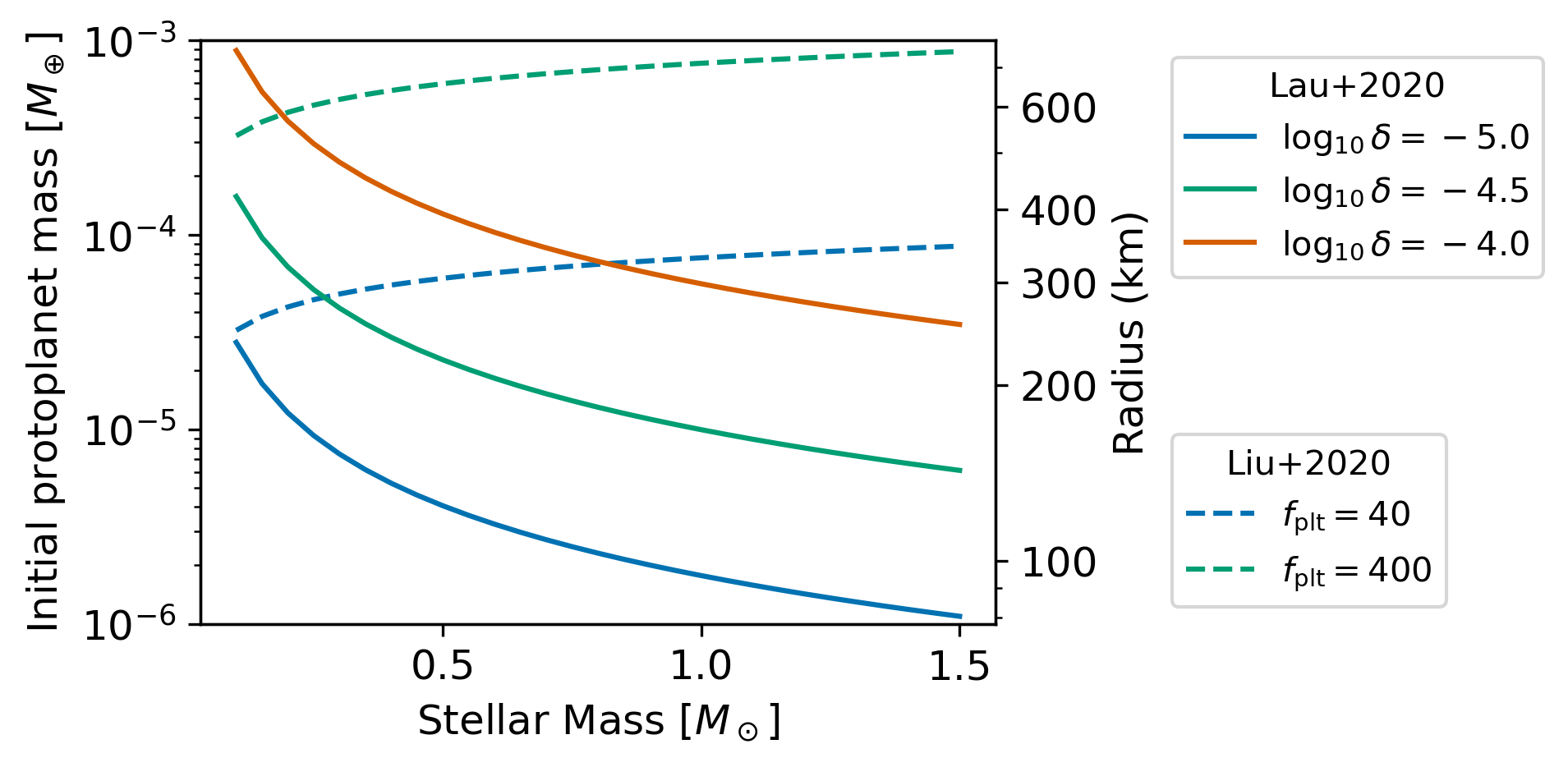}
    \caption{
    Maximum mass of planetesimals $m_\mathrm{plt}$ used in our model. The solid lines show the estimated $m_\mathrm{plt}$ in the model by \citet{Lau+2022}. The different colors correspond to different $\delta$. The dashed lines show the estimated $m_\mathrm{plt}$ in the model by \citet{Liu+2020}. The different color corresponds to different $f_\mathrm{plt}$. The right y-axis shows the radius of the protoplanet. We use the mean density of $3 \mathrm{g}/\mathrm{cm}^3$ to estimate the radius.
    }
    \label{fig: Ms_Mplt0}
\end{figure}

Hydrodynamic simulations show that the maximum size of planetesimals formed by the streaming instability could be several hundred km in radius \citep{Simon+2017, Schafer+2017}. 
We estimate the maximum size of planetesimals using two different models.
The first model is the model used in \citet{Lau+2022}, where the \revise{characteristic} planetesimal size is set by the typical clump size of pebbles determined by the local diffusivity $\delta$, which is given by \citep{Klahr+2020}: 
\begin{align}
    \revise{m_\mathrm{G}} &= \frac{1}{9} \left( \frac{\delta}{\tau_\mathrm{f}} \right)^{3/2} \left( \frac{h_\mathrm{gas}}{r} \right)^{3} M_\odot, \\
    &\sim 7.22\times10^{-3} \left( \frac{\delta/\tau_\mathrm{f}}{10^{-4}/10^{-2}} \right)^{3/2} \left( \frac{h_\mathrm{gas}/r}{0.058} \right)^{3} M_\oplus. \label{eq: mpl_Lau22}
\end{align}
\revise{
The streaming instability simulations by \citet{Abod+2019} show that the cumulative mass distribution of planetesimals exhibits an exponential cutoff $\sim0.3 m_\mathrm{G}$. Although such massive planetesimals are relatively rare, planetesimals with masses of $\sim m_\mathrm{G}$ form via streaming instability. We adopt $m_\mathrm{G}$ as the maximum planetesimal size $m_\mathrm{plt}$.}
Following \citet{Lau+2022}, we set $\delta$ as a free parameter independent of $\alpha_\mathrm{turb}$. We use $\log_{10} \delta=-4.0,-4.5$, and $-5.0$.
The second model is the model by \citet{Liu+2020}. In this model, the mass of the dust clump is determined by the self-gravity and the tidal shear, which is given by: 
\begin{align}
    m_\mathrm{plt} = 2\times10^{-3} M_\oplus \left(\frac{f_\mathrm{plt}}{400} \right) \left(\frac{\gamma}{\pi^{-1}} \right)^{1.5} \left( \frac{h_\mathrm{gas}/r}{0.05} \right)^3 \left( \frac{M_\mathrm{s}}{M_\odot} \right) \label{eq: mpl_Liu20}
\end{align}
with: 
\begin{align}
    \gamma = \frac{4 \pi \mathcal{G} \rho_\mathrm{g}}{\Omega^2_\mathrm{K}},
\end{align}
where $f_\mathrm{plt}$ is the ratio between the maximum mass and the characteristic mass of planetesimals formed by the streaming instability. \citet{Liu+2020} estimates $f_\mathrm{plt}$ as 400 from the streaming instability simulations by \citet{Schafer+2017}, but $f_\mathrm{plt}$ could change with the disk conditions \citep{Simon+2017}. To account for this uncertainty, we also use a value of $f_\mathrm{plt}=40$, which is smaller by a factor of ten. 

Figure~\ref{fig: Ms_Mplt0} shows the inferred initial mass of the protoplanet. We assume that the first planetesimal forms at $t=10^5$ yrs with $M_\mathrm{p}=m_\mathrm{plt}$. We calculate the planetary growth via pebble accretion without the gas accretion and estimate the core formation time $t_0$ by integrating it up to $10^{-2} M_\oplus$. Then, we estimate the crossover time using the numerical results obtained in Sec~\ref{sec: result1}. We introduce a fitting function $t_\mathrm{g, cross, fit} (\alpha_\mathrm{turb}, f_\mathrm{g})$ to the obtained crossover time $t_\mathrm{g, cross}$ as a function of the core formation time $t_0$ and stellar mass $M_\mathrm{s}$. 
We use \texttt{scipy.interpolate.RBFInterpolator} to determine the fitting function.

\subsubsection{Effects of initial protoplanet's mass}\label{sec: result_cross_over_time}

\begin{figure*}
    \centering
    \includegraphics[width=0.8\linewidth]{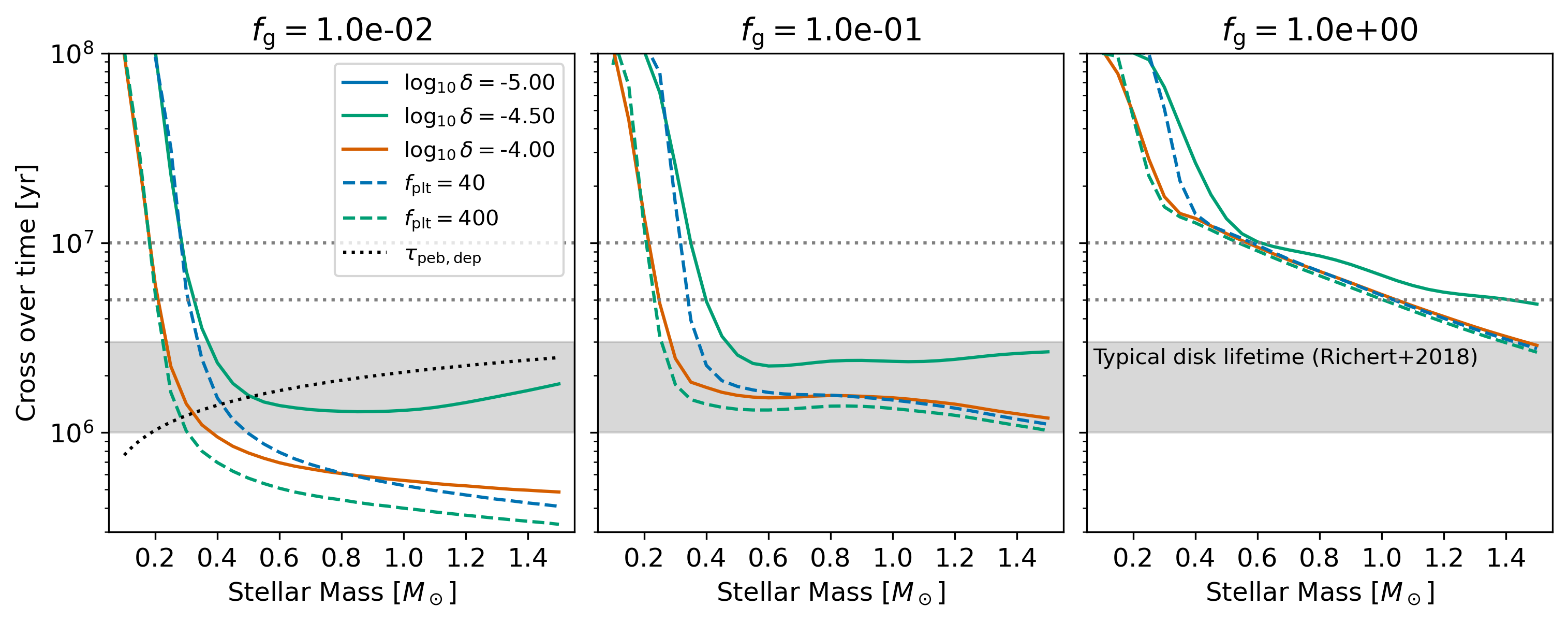}
    \caption{
    The estimated crossover time in the pebble accretion scenario. The solid and dashed lines show the cases with the planetesimal formation model by Eq.~\ref{eq: mpl_Lau22} and Eq.~\ref{eq: mpl_Liu20}, respectively. The different colors show the cases with different parameters, as described in the legend box. Each panel shows the cases with different grain opacities. The gray area shows the typical disk lifetime inferred by the observations \citep{Richert+2018}, which ranges from 1 Myrs to 3 Myrs. We plot the horizontal gray dotted lines at $5$ Myrs and $10$ Myrs for the eye guide.
    }
    \label{fig: t_cross_over_pskap}
\end{figure*}

\begin{figure}
    \centering
    \includegraphics[width=0.8\linewidth]{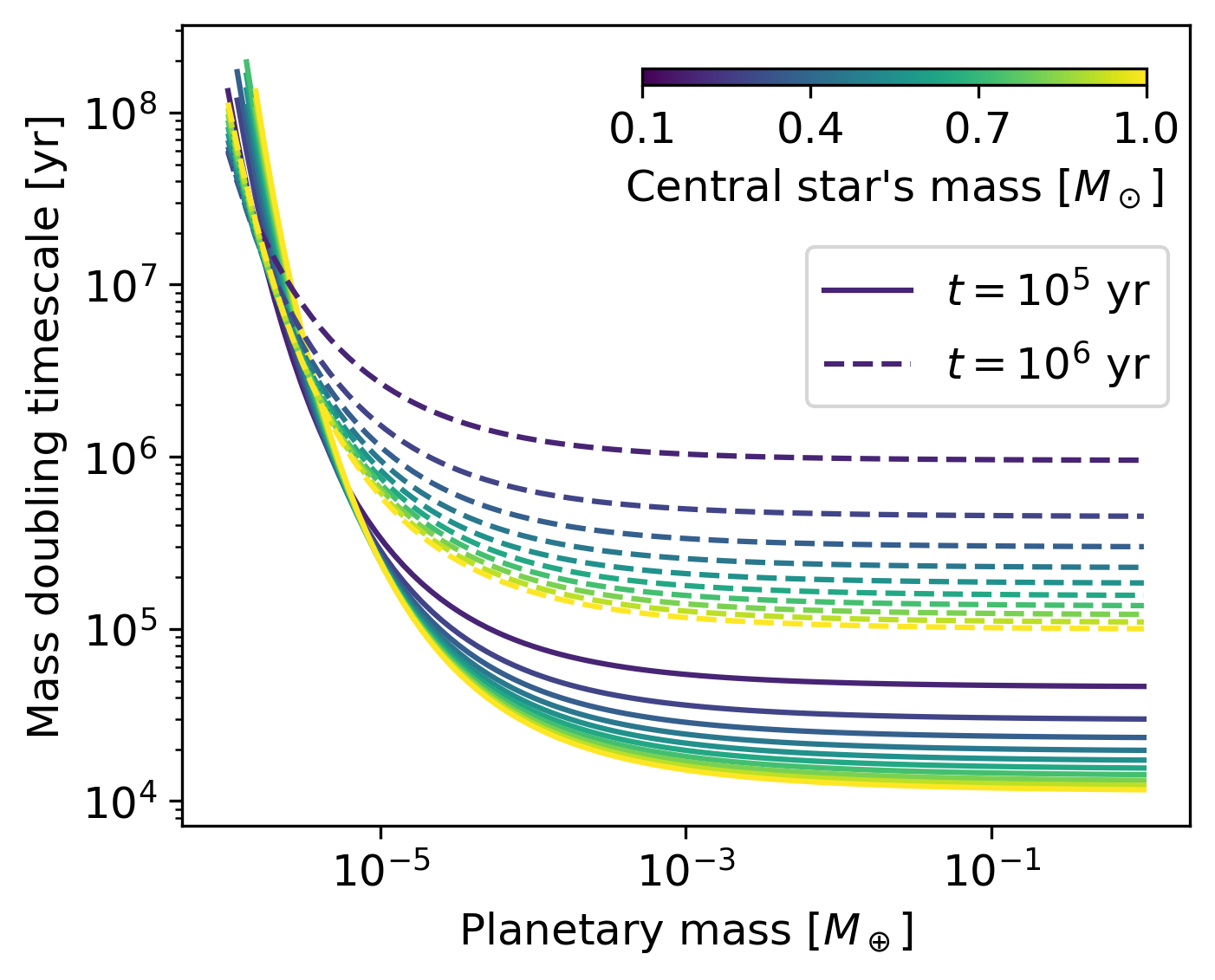}
    \caption{
    Mass doubling timescale in the pebble accretion model. The solid and dashed lines show the different times $t=10^5$ yrs and $10^6$ yrs, respectively. The color corresponds to the different masses of central stars, and the color bar is written on the panel's upper right.
    }
    \label{fig: pebble_gtime}
\end{figure}

Figure~\ref{fig: t_cross_over_pskap} shows the estimated crossover time. We plot the cases using different initial protoplanet masses $m_\mathrm{plt}$. First, the protoplanet could not reach the crossover mass when we use Eq.~\ref{eq: mpl_Liu20} with $\delta=10^{-5}$. In this case, the mass of the initial protoplanet is $\lesssim10^{-5} M_\oplus$ and too small to initiate rapid pebble accretion. We present the mass-doubling timescale $\tau_\mathrm{peb}$ in Fig.~\ref{fig: pebble_gtime}. As shown in this figure, the pebble accretion rate strongly depends on the mass of the growing protoplanet when $\lesssim10^{-5} M_\oplus$. To initiate rapid pebble accretion, the initial protoplanet's mass must be larger than $\gtrsim10^{-5} M_\oplus$.

The protoplanet can grow to the crossover mass with the other initial protoplanet mass. For  $f_\mathrm{g}=1$, we find that the crossover time rarely depends on the initial protoplanet mass. This is because the growth time in the contraction phase is longer than in the core growth phase.

When the grain opacity factor is small as $f_\mathrm{g}=0.01$, the crossover time is shorter for the larger initial protoplanet mass. However, even if the initial mass was 10 times larger, the crossover time is shortened by 2-3. In the pebble accretion model, the mass-doubling timescale rarely changes with the protoplanet mass when $M_\mathrm{p}\gtrsim10^{-5} M_\oplus$ as shown in Fig.~\ref{fig: pebble_gtime}. If the mass doubling timescale $\tau_\mathrm{peb}$ is constant, the protoplanet mass is written as
\begin{align}
    \frac{M_\mathrm{p}}{\dot{M}_\mathrm{p}} &= \tau_\mathrm{peb}, \nonumber \\
    \log \frac{M_\mathrm{p}}{m_\mathrm{plt}} &= \frac{t}{\tau_\mathrm{peb}}.
\end{align}
Since the contraction phase is negligibly shorter than the core growth phase, the time reaching the crossover is given by 
\begin{align}\label{eq: crossover_core_growth_regime}
    t_\mathrm{g, cross} = \tau_\mathrm{peb} \left(\ln M_\mathrm{iso} - \ln m_\mathrm{plt}\right).
\end{align}
Therefore, the crossover time weakly depends on the initial protoplanet mass if the initial planetesimals are born big as $\gtrsim10^{-5} M_\oplus$.

From the above results, we conclude that the protoplanet's initial size has a minor effect on the crossover time in the pebble accretion scenario, compared to the other parameters, like the grain opacity and stellar mass.

\subsubsection{Crossover time across different stellar types}

We find that the crossover time decreases with increasing stellar mass and changes the dependence on the stellar mass around $M_\mathrm{s}~\sim~0.4 M_\odot$ as shown in Fig.~\ref{fig: t_cross_over_pskap}. For planets orbiting around stars of $M_\mathrm{s}~\lesssim~0.4 M_\odot$, the crossover time strongly depends on $M_\mathrm{s}$. These planets could not reach pebble isolation mass before the pebbles in the protoplanetary disk are depleted. The pebble flux is generated by the sweep of the pebble growth line $r_\mathrm{g}$. When $r_\mathrm{g} < R_\mathrm{disk}$, the pebble flux slowly decreases with time ($\propto t^{-8/21}$). Once the pebble growth line reaches the disk's typical radius $R_\mathrm{disk}$, the pebble flux decreases exponentially because of the reduction in the dust surface density at the disk's outer edge. By substituting $r_\mathrm{g} = 2 R_\mathrm{disk}$ into Eq.~\ref{eq: pebble_production_line}, $r=2 R_\mathrm{disk}$ gives a ten times reduction in $\Sigma_\mathrm{gas}$ because of the disk's outer edge, we obtained the pebble depletion time as: 
\begin{align}\label{eq: t_peb_dep}
    \tau_\mathrm{peb, dep} = 2.08 \mathrm{Myr} \left( \frac{M_\mathrm{s}}{1 M_\odot} \right)^{\frac{3}{2} BC - \frac{1}{2}} \left( \frac{\epsilon_\mathrm{D}}{0.05} \right)^{-1} \left( \frac{Z_\mathrm{disk}}{0.01} \right)^{-1}.
\end{align}
We plot Eq.~\ref{eq: t_peb_dep} on the left panel in Fig.~\ref{fig: t_cross_over_pskap}. 

When $f_\mathrm{g}=0.01$, the envelope contraction phase is much shorter than the core growth phase, as we found in Sec.~\ref{sec: result1}. $t_\mathrm{g, cross}<\tau_\mathrm{peb, dep}$ means that the protoplanet can reach the pebble isolation mass before the pebbles are depleted. The dependence of the crossover time on the stellar mass comes from the core growth time, which decreases with increasing stellar mass because the mass doubling timescale $\tau_\mathrm{peb}$ is shorter around the higher mass stars.  
On the other hand, the protoplanets could not reach the pebble isolation mass in the cases with $t_\mathrm{g, cross}>\tau_\mathrm{peb, dep}$. For these non-isolated protoplanets, the envelope's contraction takes a longer time because of the smaller core mass. The protoplanet's core mass is smaller around the smaller mass stars because of the shorter $\tau_\mathrm{peb, dep}$ and longer $\tau_\mathrm{peb}$. Therefore, the crossover time steeply increases with decreasing stellar mass. 

We plot the typical disk lifetime $1-3$ Myrs with the gray area in Fig.~\ref{fig: t_cross_over_pskap}. Using the dust observations in stellar clusters, \citet{Richert+2018} found that the disk longevity rarely depends on the stellar mass, and the fraction of stars with disks is $60$-$70\%$ at 1 Myrs and $30$-$40\%$ at 3 Myrs. 
\revise{
Currently, there is no evidence that the lifetimes of the dust disk and the gas disk are the same. However, it is reasonable to assume that both components could be dispersed by the same mechanisms, such as photoevaporation or disk winds.  However, dust growth could lead to the selective removal of small dust grains while leaving the gas behind, and possibly leading to a gaseous disk that survives longer than dust disks inferred from observations. For simplicity, in this study the lifetimes of the dust and gaseous disks are assumed to be equal. Under this assumptions our results suggest 
} 
that more than $30\%$ of stars heavier than $\sim0.4 M_\odot$ have cold Jupiters if $f_\mathrm{g}\lesssim0.1$. This result is inconsistent with the observed occurrence rate of cold Jupiters \citep{Fulton+2021}, where only $\lesssim20\%$ of G-type stars have cold Jupiters.

To be consistent with the low occurrence rate of cold Jupiters, a higher grain opacity of \revise{$f_\mathrm{g}>0.1$} is required. When $f_\mathrm{g}=1$, the crossover time ranges between $5$-$10$ Myrs in \revise{$M_\mathrm{s}\gtrsim0.4 M_\odot$}. Also, the crossover time decreases with the stellar mass. Due to the high grain opacity, the crossover time is regulated by the envelope's contraction. The timescale in Kelvin-Helmholtz contraction is given by \citep{Ikoma+2000}: 
\begin{align}\label{eq: timescale_KH}
    \tau_\mathrm{KH} = 10^6 \mathrm{yr} \left( \frac{M_\mathrm{p}}{10 M_\oplus} \right)^{-k_2} \left( \frac{\kappa_\mathrm{env}}{1 \mathrm{cm^2}/\mathrm{g}} \right),
\end{align}
where $\kappa_\mathrm{env}$ is the effective opacity in the planetary atmosphere and $k_2$ is estimated $2.5-3.5$. In our model, the pebble isolation mass scales as: 
\begin{align}\label{eq: M_iso_model}
    M_\mathrm{iso} \sim 8 M_\oplus \left( \frac{M_\mathrm{s}}{M_\odot} \right)^{\frac{3}{7}}.
\end{align}
Using $k_2=3.0$ and substituting $M_\mathrm{p}=M_\mathrm{iso}$, we get $\tau_\mathrm{KH} \propto M^{-9/7}_\mathrm{s}$. The strong dependency between the   crossover time and the stellar mass could explain the higher occurrence rate of cold Jupiter around more massive stars, as inferred from observations \citep{Fulton+2021}. 

The exact occurrence rate also depends on the distribution of the disk's lifetime. However, the available observational data are limited to $t\leq5$ Myrs in \citet{Richert+2018}, and the fraction of stars with disks between $5$-$10$ Myrs is unclear yet. Recent studies suggest that the characteristic timescale for the disk's lifetime could be $5-10$ Myr \citep{Michel+2021, Pfalzner+2022}, which prefers the scenario with the high grain opacity $f_\mathrm{g}=1$ to be consistent with the observed cold Jupiter's occurrence rate. Since the formation of giant planets is controlled by the gaseous disk's lifetime, a better understanding of the distribution of disk lifetimes is required. 

\revise{
Also, the gas distribution in the protoplanetary disks was assumed to be smooth. However, this is a clear simplification. Indeed, recent observations of protoplanetary disks show the common existence of substructures such as rings and spirals \citep[e.g.][]{Andrews2020, Benisty+2022}. These substructures would affect the pebble flux and its accretion onto protoplanets. For example, \citet{Lau+2022} shows that if planetesimals form and grow in a pressure bump, planetesimal accretion and rapid pebble accretion can occur, leading to the formation of a core of several $M_{\oplus}$ within $10^5$ years. If such substructures shorten the core growth phase, it emphasizes the need for a slow contraction phase. It is therefore clear that a better understanding of the mechanisms by which substructures in protoplanetary disks accelerate or delay the formation of giant planets is required. This will allow us to better connect the properties of the planetary disk and the planetary formation history. 
}


\section{Discussion \& caveats}\label{sec: discussion}

\subsection{Heavy-element deposition in the envelope}\label{sec: discussion_deposition}

\begin{figure}
    \centering
    \includegraphics[width=1\linewidth]{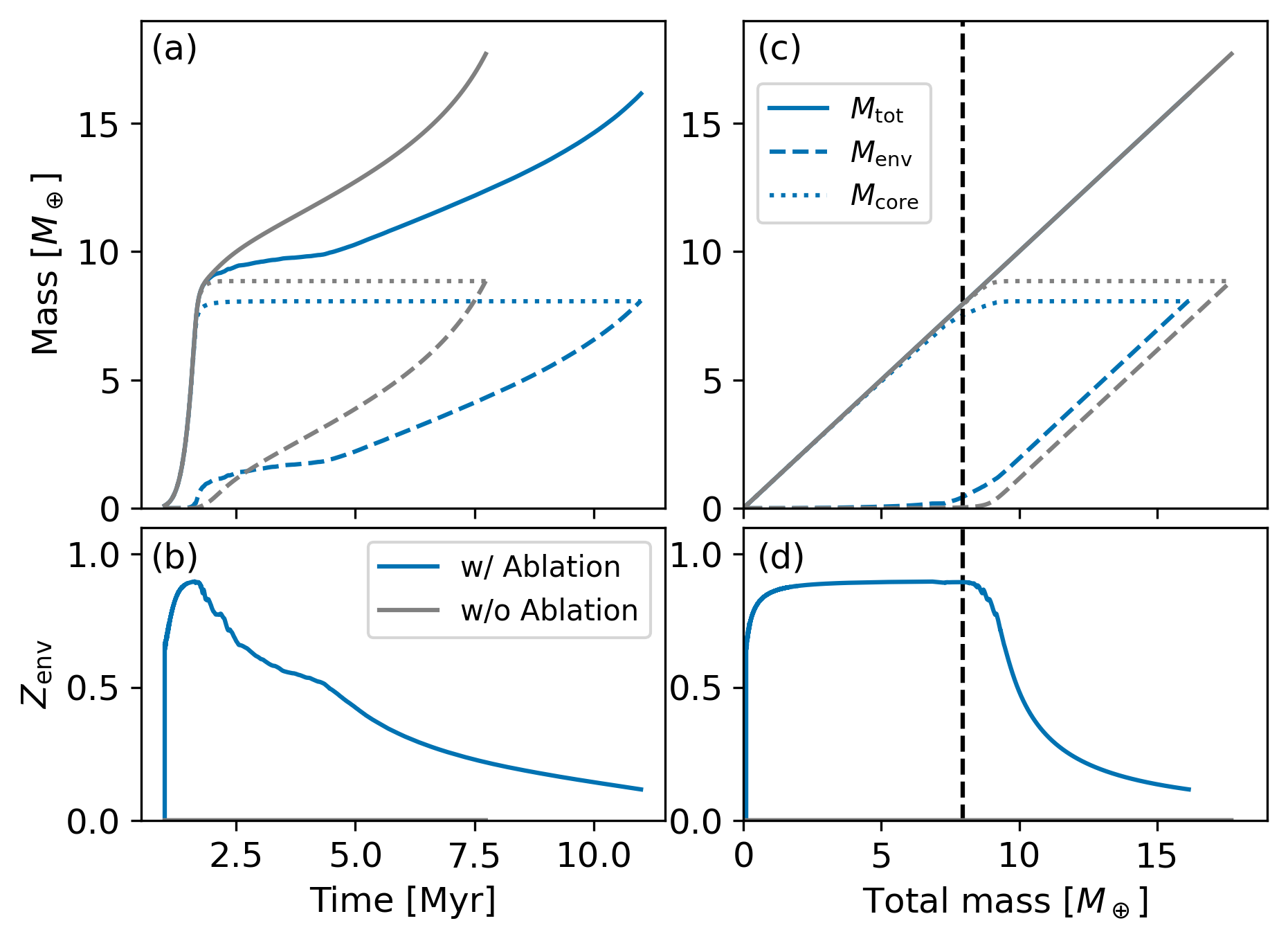}
    \caption{
    The growth of the planetary core up to the crossover mass accounting for envelope enrichment caused by pebble evaporation. The blue lines show the case with $M_\mathrm{s}=1M_\odot$, $t_0=10^6\mathrm{yrs}$, $\alpha_\mathrm{turb}=10^{-3}$ and $f_\mathrm{g}=1$. The gray lines show the case without envelope enrichment.
    {\bf Panel-(a)}: Time evolution of total mass (solid line), envelope mass (dashed line), and core mass (dotted line). {\bf Panel-(b)}: Time evolution of the bulk metallicity of planetary envelope. {\bf Panel-(c)}: Evolution of planetary masses as a function of the total mass. {\bf Panel-(b)}: Evolution of the bulk metallicity of planetary envelope vs.~the total mass of protoplanet. 
    The vertical dashed lines in Panel-(c) and (d) are pebble isolation mass.
    }
    \label{fig: grow_enrichEnv}
\end{figure}

\begin{figure}
    \centering
    \includegraphics[width=1\linewidth]{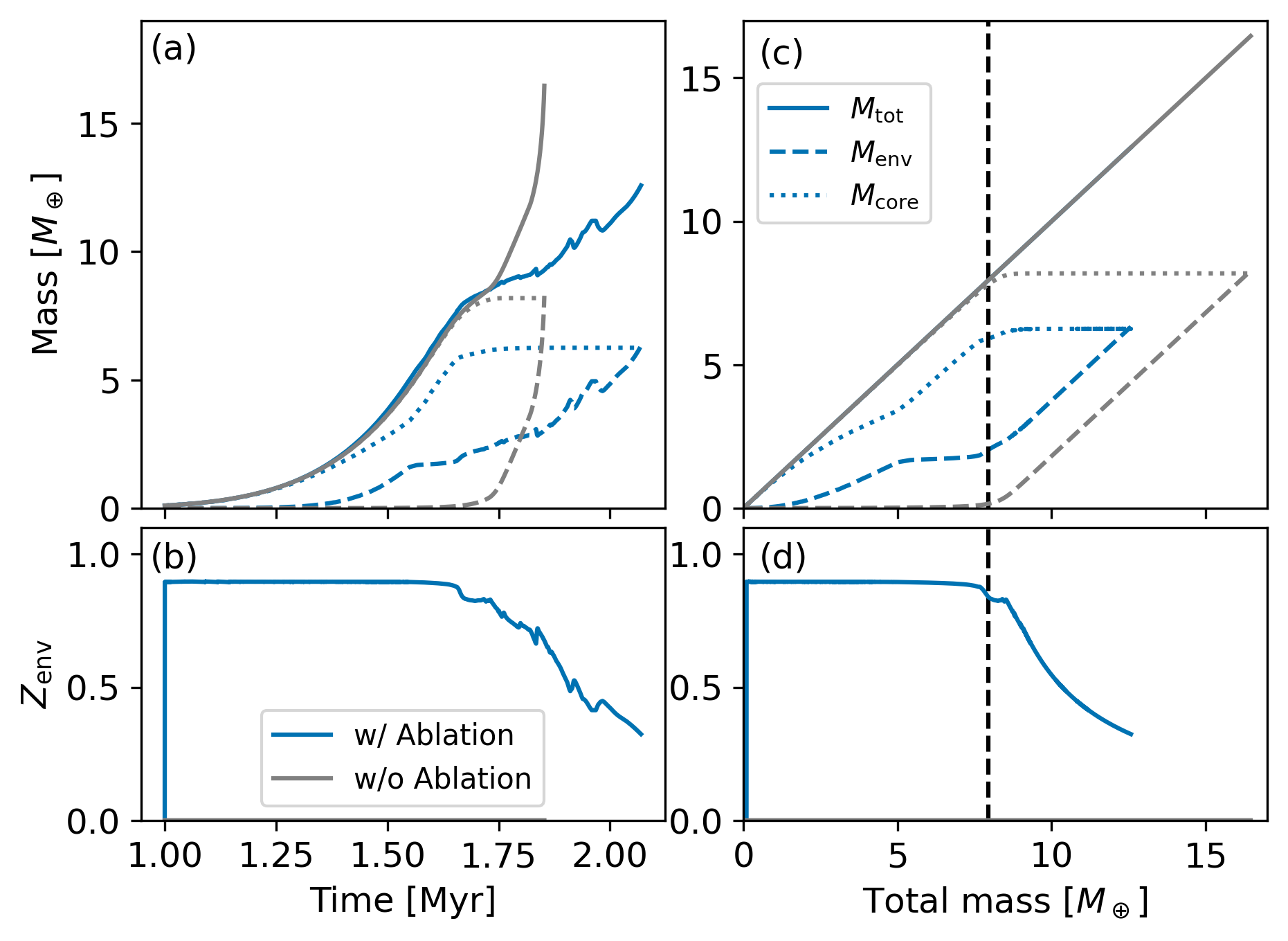}
    \caption{
    Same as Fig~\ref{fig: grow_enrichEnv}, but with $f_\mathrm{g}=0.01$.
    }
    \label{fig: grow_enrichEnv_fg001}
\end{figure}

In the simulations presented above, we neglect the ablation of the accreted pebbles and assume that all the heavy elements settle to the center. However, the enrichment of the envelope with heavy elements accelerates the contraction of the planetary envelope \citep{Hori+2010, Hori+2011, Venturini+2015, Venturini+2016, Valletta+2020, Mol-Lous+2024}. In this section, we investigate how envelope enrichment changes the crossover time. 

We use the ablation model developed by \citet{Valletta+2020} and \citet{Mol-Lous+2024}. We use the direct deposition model, where the ablated heavy elements from the accreted pebbles are deposited into the local envelope shell. We set the pebble composition to be 50 \% rock and 50 \% H$_2$O, and assume that only the ablated H$_2$O is deposited in the envelope. We include the condensation of H$_2$O using the saturation pressure and the maximum water enhancement of the supercritical water $Z_\mathrm{sup, H_2O}=0.9$. 
We also introduce the pre-mixing algorithm for heavy elements if the Rayleigh-Taylor criterion is met. Further details of our heavy-element enrichment method are presented in Appendix~\ref {sec: method_Z}. To save computational time, we start with $M_\mathrm{0}=0.1 M_\oplus$, instead of $M_\mathrm{0}=0.01 M_\oplus$. 
%

Figure~\ref{fig: grow_enrichEnv} shows the planetary evolution including the envelope enrichment by the accreted pebbles. We find that the metallicity in the protoplanet envelope increases quickly and saturates to $\sim0.8$. After reaching the pebble isolation mass, the atmospheric metallicity decreases because only gas accretes to the protoplanet in this phase. We also plot the cases without envelope enrichment with gray lines. Note that these cases differ from the simulations in Sec.~\ref{sec: result1} because of different $M_\mathrm{0}$. Interestingly, we find that the crossover time is longer for the cases with envelope enrichment. The high metallicity of the envelope leads to higher gas accretion rates and larger envelope mass during the core growth phase. Since the total protoplanet mass regulates the pebble isolation mass, the core mass decreases at the pebble isolation compared to the cases without envelope enrichment.  \citet{Piso+2014} shows that the gas accretion rate in the following contraction phase strongly depends on the core mass. Due to the smaller core, the growth time in the contraction phase is longer when envelope enrichment is considered. This situation does not occur in the study of \citet{Mol-Lous+2024} because the pebble isolation mass was set to $\sim 25 M_\oplus$ and the protoplanet did not reach the pebble isolation in 10 Myrs. 

We also performed simulations with a lower grain opacity factor $f_\mathrm{g}=0.01$ (Fig.~\ref{fig: grow_enrichEnv_fg001}). Due to the rapid envelope contraction in the contraction phase, the effect of envelope enrichment is weaker. The inferred difference in the crossover time is only $\lesssim0.3$ Myrs. 

We find that heavy-element enrichment in the planetary envelopes does not necessarily shorten the formation timescale of giant planets. However, we investigated only a few parameter studies using the specific enrichment parameters (50\%-rock-50\%-water and $Z_\mathrm{sup,H_2O}=0.9$). As shown in \citet{Mol-Lous+2024}, the effect of envelope enrichment is more profound if the accreted pebbles are composed of pure-water because the energy deposited by the accreted pebbles is small. Even with 50\%-rock-50\%-water, the ablation and deposition of rock, which we assume non-missible to the H-He envelope, would accelerate gas accretion and would change the mass of the planetary core. Future studies should investigate in further detail how the planetary formation timescale changes with envelope enrichment when considering various compositions.

\subsection{Extending the envelope contraction phase}\label{sec: discussion_envelope_contraction}

We find that a high grain opacity factor of \revise{$f_\mathrm{g}>0.1$} is required to explain the low occurrence rate of cold Jupiters. However, dust growth models in planetary envelopes suggest that small dust grains grow to larger sizes and quickly settle to deeper regions, reducing grain opacity \citep{Movshovitz+2010, Mordasini2014, Ormel2014}. The estimated grain opacity factor is $f_\mathrm{g}\leq0.01$ \citep{Mordasini2014}. 
Additional mechanisms that  delay the envelope's contraction, such as gas recycling or a continuous accretion of dust and planetesimals after the protoplanet reached pebble isolation mass are required. 

3D hydrodynamical simulations show that the accreted gas enters the Hill sphere through the poles and exits to the disk's midplane \citep{Tanigawa+2012, Ormel+2015, Kurokawa+2018}. The recycled gas keeps bringing small dust to the radiative-advective boundary region of the planets, which would lead to a dust opacity comparable to the (local) dust opacity in the disk \citep{Lambrechts+2017}. After pebble isolation mass is reached, the drifting pebbles pile up at the outer pressure bump. This can trigger fragmentation, and dust grains that are small enough to be coupled to the disk gas can flow into the planetary envelope \citep{Stammler+2023}. The accretion of small dust could increase the grain opacity in the planetary envelope and may delay the envelope's contraction \citep{Chen+2020}. Further investigation of the grain opacity in protoplanetary disks, including the grain growth around a protoplanet, should be conducted in future studies.  

Another possible mechanism is planetesimal accretion. 
\revise{
In this study we neglected planetesimal accretion after the formation of initial planetesimals. No protoplanets grow via pebble accretion if the initial protoplanet mass $m_\mathrm{plt}$ is smaller than $10^{-5} M_\oplus$. However, planetesimal accretion could grow the small planetesimals to a point where they trigger rapid pebble accretion \citep[e.g.,][]{Johansen+2017}. The timing of when rapid pebble accretion starts could be later than the time of planetesimal formation, and the crossover time could be more delayed than that obtained in our simulations. 
Also,}
we neglected the process of planetesimal accretion which could take place in parallel to pebble accretion. Unlike pebble accretion, planetesimal accretion could continue even after the gap formation, and accreted planetesimals could provide additional energy to the planetary envelope, delaying the contraction of the planetary envelope \citep{Alibert+2018, Helled2023}. Since planetesimals need to form to start rapid pebble accretion, planetesimal accretion seems to be inevitable during giant planet formation, although the accretion rate could be smaller than that of pebble accretion. In fact, it is possible that pebble accreton and planetesimal accretion dominate at different stages during planet formation as well as in different disk environments. Understanding the interplay between planetesimal accretion and pebble accretion is desirable.  






\subsection{Effects of disk scaling models}
The pebble accretion process strongly depends on the assumed  disk model. In our disk model, protoplanets could not reach pebble isolation around stars of $\lesssim0.4M_\odot$, and the break point of the crossover time where the dependency on stellar mass changes appears around $0.4 M_\odot$. However, as shown in Eq.~\ref{eq: t_peb_dep}, the pebble depletion time decreases around low-mass stars, and the break point would shift to larger mass if the exponent $BC$ gets larger. 

In our model, the disk accretion rate $\dot{M}_\mathrm{d, acc}$ scales as:
\begin{align}
    \frac{\mathrm{d} \log \dot{M}_\mathrm{d, acc} }{\mathrm{d} \log M_\mathrm{s}} \sim {\frac{2}{7} A +B \left\{ 1- C \left( 2 - \gamma \right) \right\} } = \frac{39}{112}.
\end{align}
Here, we neglect the viscous evolution and radial decay terms in $\Sigma_\mathrm{gas}$. The relation between the disk accretion rate and the stellar mass can be estimated from observations \citep[e.g.][]{Manara+ppii}, and our scaling law seems to be shallower than that in observations. Single power-law fitting suggests a slope of $1.6-2.$. 
Such a steep slope requires higher $A$, $B$, or smaller $C$ than those in our model. However, the exact relation between $\dot{M}_\mathrm{d, acc}$ and $M_\odot$ is still actively discussed. Several studies show that a double power-law fit for $\dot{M}_\mathrm{d, acc}-M_\odot$ is a more apropriate representation than a simple power-law fit \citep{Alcala+2017, Manara+2017}. In this case, the slope is flatter for stars more massive than $0.2-0.3 M_\odot$. 
Note that the observed disk accretion rate is time-dependent and that the exact scaling law at time zero remains unknown. 
A more robust determination of the properties of disks around different stellar types is important for constraining the origin of cold Jupiters and predicting their occurrence rate.

\revise{
\subsection{Planetary migration}\label{sec: discussion_migration}
}

\revise{
For simplicity, planetary migration is not included in this study. However, it is clear that planet migration plays an important role in planet formation. The migration regime typically changes from type I to type II around the pebble isolation mass ($\sim 10 M_\oplus$), and the migration speed peaks around this transition \citep{Kanagawa+2018, Ida+2018}. When the grain opacity is as high as $f_\mathrm{g}=1$, the protoplanet enters a long contraction phase after reaching pebble isolation mass. During this phase, the protoplanet can undergo significant radial migration, potentially becoming a warm or hot Jupiter. In contrast, when the grain opacity is low, e.g., $f_\mathrm{g}=0.01$, the protoplanet grows rapidly after pebble isolation mass is reached  (see Fig.~\ref{fig: time_growth_psfg}) and quickly shifts to the  slower type II migration. This quick transition to slower migration could help prevent significant inward migration. Note, however, that it remains unclear whether  such low grain opacities are common. 
}

\revise{
While inward migration is expected to occur,  outward migration \citep{Paardekooper+2010, Paardekooper+2011, Paardekooper+2014}, and migration trap \citep{Masset+2006, Coleman+2016} could also take place. 
Outward migration occurs near the transition radius, where the dominant disk heating mechanism shifts from viscous accretion to stellar irradiation \citep{Liu+2019, Liu+2020}. However, this mechanism weakens as the planet grows in mass and the disk evolves. Alternative mechanisms are disk substructures that can act as migration traps, enabling the survival of cold Jupiters \citep{Coleman+2016}. Both outward migration and migration traps are sensitive to the evolving disk structure. A more comprehensive understanding of the formation of "cold" giant planets will require future investigations that incorporate planetary migration self-consistently, ideally through disk evolution models and hydrodynamical simulations.
}

\revise{
If significant planetary migration occurs, quick transition to slower type II migration might be required. To be consistent with the lower occurrence rate of cold Jupiters, planetary cores should form less efficiently. Our model assumes that icy pebbles are sticky and continue to grow until the gas drag from the disk causes the growing pebbles to drift inward. However, recent observations of dust in protoplanetary disks \citep{Jiang+2024} and laboratory experiments \citep{Musiolik+2016} suggest that icy pebbles are more fragile and pebbles initiate fragmentation even outside the water ice line. If this is the case, the pebble accretion efficiency is reduced due to the lower Stokes number. Since the planetary core growth takes longer, giant planets would form less efficiently, even if the grain opacity is lower than $f_\mathrm{g}=1$. Fragile icy pebbles may support the formation of cold Jupiters.
}

\revise{
It remains unclear how planet migration takes place in disks around different stellar types and we are still lacking a good understanding of how the disk properties (e.g., density, lifetime) change with the mass of the central star. As a result, it is still unclear whether different migration histories are expected. Future studies should investigate how planetary migration affects planet formation across different stellar masses.  
}


\section{Summary and conclusions}\label{sec: conclusion}

\begin{figure*}[h!]
    \centering
    \includegraphics[width=0.8\linewidth]{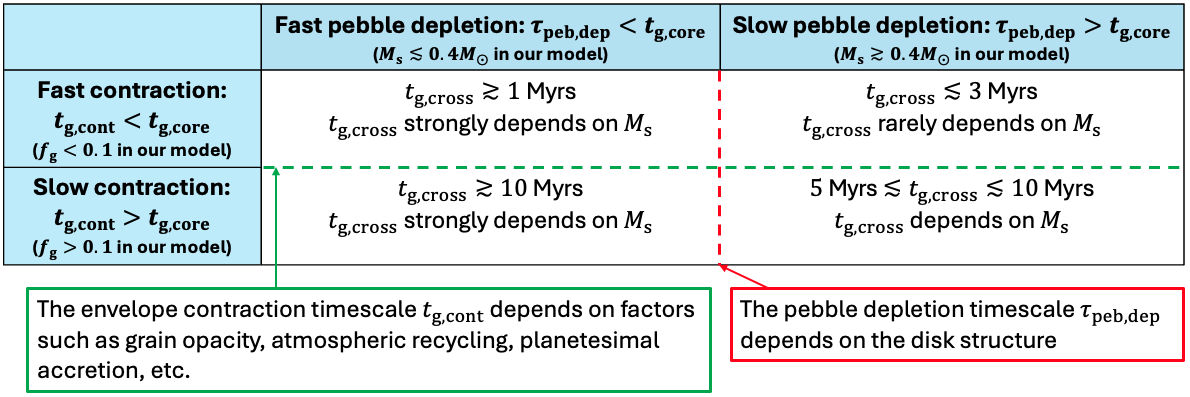}
    \caption{
    A summary of our results and the inferred giant planet formation regimes. 
    }
    \label{fig: summary}
\end{figure*}

We investigated the formation of cold Jupiters around different mass stars. We explored how the crossover time changes with stellar mass $M_\mathrm{s}$, core formation time $t_0$, grain opacity factor $f_\mathrm{g}$, and disk turbulent viscosity $\alpha_\mathrm{turb}$, assuming pebble accretion. 
Our key conclusions can be summarized as follows: 
\begin{itemize}
    \item The grain opacity strongly affects the planetary growth. If the grain opacity is lower than $f_\mathrm{g}\lesssim0.1$, the crossover time is comparable to the growth timescale of the heavy-element core. Contrary,  if $f_\mathrm{g}\gtrsim0.1$, the envelope contraction dominates the formation process, and the protoplanets experience a long envelope contraction phase after reaching the pebble isolation mass.
    \item Low disk turbulent viscosity $\alpha_\mathrm{turb}$ accelerates (delays) reaching crossover mass when the grain opacity is lower (higher) than $f_\mathrm{g}\sim0.1$. When $f_\mathrm{g}<0.1$, the core growth time decreases for lower $\alpha_\mathrm{turb}$ because of the smaller \revise{pebble layer's} scale height and higher pebble accretion rate. On the other hand, the envelope contraction timescale increases because of the lower pebble isolation mass if $f_\mathrm{g}>0.1$. 
    \item The crossover time depends on the initial size of protoplanets. However, this dependency is weaker than linear. The grain opacity plays a bigger role than the initial size of protoplanets in determining the crossover time.
    \item Around stars of $M_\mathrm{s}\lesssim0.4M_\odot$, the crossover time rapidly increases with decreasing stellar mass. Around these stars, pebbles in the protoplanetary disk deplete before the protoplanets reach pebble isolation mass. Because of the small mass of the forming solid core, the envelope's contraction timescale (before reaching the crossover mass) is long. A very long disk lifetime $\gtrsim10^7$ yrs, as found in Peter Pan disks, is required to form giant planets around such stars.
    \item Around stars with masses \revise{$M_\mathrm{s}\gtrsim0.4M_\odot$}, protoplanets reach crossover mass within $3$ Myrs if the grain opacity is  $f_\mathrm{g}\lesssim0.1$, which is comparable or shorter than the typical disk's lifetime. To be consistent with the low occurrence rate of cold Jupiters around G-type stars ($\lesssim20\%$), a high grain opacity  ($f_\mathrm{g}>0.1$) is required. With a high grain opacity $f_\mathrm{g}=1$, the crossover timescale increases around lower-mass stars, which could explain the inferred low-occurrence rate of cold Jupiters around small stars. 
\end{itemize}

We find that the crossover time can be divided into two regimes by comparing the core formation time $t_\mathrm{g,core}$ and the pebble depletion timescale $\tau_\mathrm{peb,dep}$. In the fast pebble depletion regime where $\tau_\mathrm{peb,dep}<t_\mathrm{g,core}$, the protoplanet does not reach the pebble isolation mass, and the crossover time 
\revise{
exceeds 10 Myrs because of the small core mass. Giant planets may form in a long-lived disk, such as the Peter-Pan disk, but this formation path would not be the primary formation path of cold Jupiters around FGK-type of stars.
}

\revise{
If the planetary core forms before the pebbles are depleted (slow pebble depletion regime where $\tau_\mathrm{peb, dep}>t_\mathrm{g, core}$), the protoplanet can grow to the pebble isolation mass. To account for the observed low occurrence rate of cold Jupiters around G-type stars, either slow envelope contraction or slow core formation—lasting longer than the typical disk lifetime—is required. In our pebble accretion model, protoplanets quickly reach the pebble isolation mass and therefore, a slow contraction of the planetary envelope is required to suppress giant planet formation and explain the low occurrence rate of cold Jupiters. 
}
This result is consistent with recent formation models that suggest that the formation of Jupiter and Saturn could have taken a few million years \citep{Alibert+2018, Helled2023}. However, rapid contraction may be required to form giant planets around lower-mass stars. This result may indicate that the metallicity and dust content of protoplanetary disks increase with the mass of the central stars. Indeed, it was recently shown that the bulk metallicity of giant planets is lower for giant planets orbiting small stars in comparison to giant planets around sun-like stars \citep{Muller+2025}. The relationship between the heavy-element mass in the disk and stellar mass is yet to be determined, and is likely to control giant planet formation around different stellar masses. 

Interestingly, we found that envelope enrichment with heavy elements released from accreted pebbles could delay envelope contraction, a different result from that obtained in previous studies. Because of the higher mean molecular weight, the protoplanets could have a more massive envelope at the pebble isolation mass. 
\revise{Consequently, the core mass at the pebble isolation is smaller when envelope enrichment is considered.}
Due to the smaller core mass, the envelope contracts more slowly, resulting in a delayed crossover time. 
\revise{
Our results also do not rule out the possibility of slow core formation. The timescale for core growth depends sensitively on disk structure, metallicity, and the dust fragmentation velocity.
}

\revise{
We note that the durations of core formation and envelope contraction depend on the formation history and therefore on various processes and parameters. The core formation timescale strongly depends on the local conditions in the disk and the types of solids that are accreted. Similarly, while in this study, a slow contraction phase is achieved with a high grain opacity, delaying the  envelope contraction could be the result of other mechanisms such as atmospheric recycling or planetesimal accretion. It is therefore clear that a better understanding of the conditions and processes taking place during planet formation is crucial. 
}
\revise{
Overall, our findings highlight the complex interplay between core accretion, envelope physics, and disk evolution in shaping planetary outcomes. 
}
We suggest that future studies should include these processes self-consistently and investigate a wide parameter space. Only then can we reveal the origin of cold Jupiters in our galaxy.

\begin{acknowledgements}
The numerical computations were carried out on PC cluster at the Center for Computational Astrophysics, National Astronomical Observatory of Japan. RH acknowledges support from SNSF under grant \texttt{\detokenize{200020_215634}}. 
\end{acknowledgements}

%
\bibliographystyle{aa} 
\bibliography{ref} 
%

\begin{appendix} 

\section{Solid mass conservation}\label{app: disk_model}

\begin{figure}
    \centering
    \includegraphics[width=1.\linewidth]{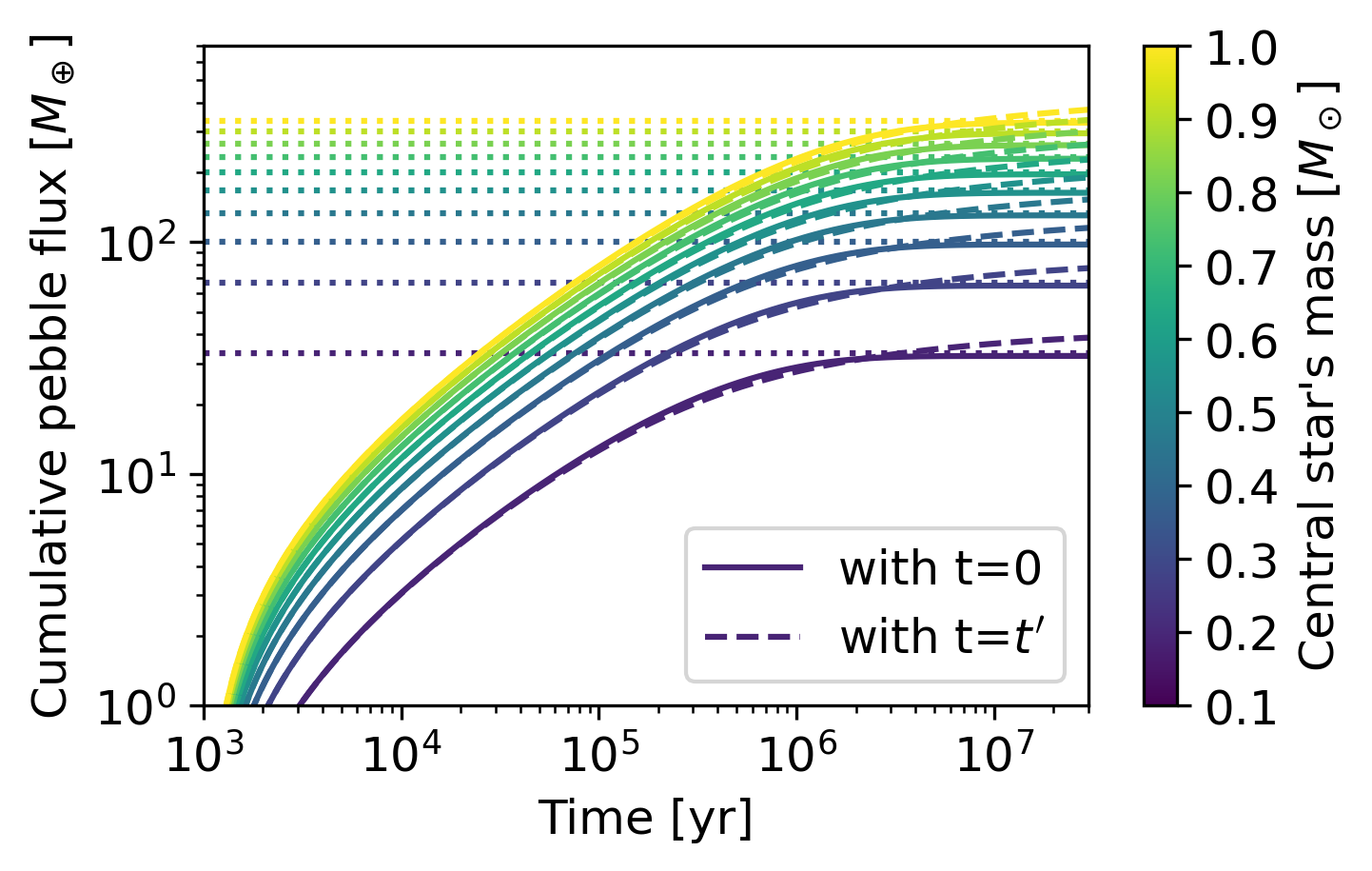}
    \caption{
    \revise{
    Time evolution of the cumulative pebble flux. The solid lines show the cumulative pebble flux calculated with Eq.~\ref{eq: peb_flux}. The dashed lines are those obtained with Eq.~\ref{eq: peb_flux} using the time evolving $\Sigma_\mathrm{gas}(t=t^\prime)$ instead of $\Sigma_\mathrm{gas}(t=0)$. The horizontal dotted lines show the initial total mass of solids in the protoplanetary disk. The different colors correspond to disks around different stellar masses. 
    }
    }
    \label{fig: method_disk_model}
\end{figure}

\revise{
In Eq.~\ref{eq: peb_flux} we use the solid surface density at $t=0$ instead of that at $t=t^\prime$, where $t^\prime$ is the disk's age.  
Using $\Sigma_\mathrm{gas} (t=t^\prime)$ in Eq.~\ref{eq: peb_flux} would account for the radial transport of small dust grains due to gas advection and diffusion, and thus seems more realistic than using $\Sigma_\mathrm{gas} (t=0)$. However, this approach overestimates the pebble flux. This is because the gas that diffuses outward from $r<R_\mathrm{disk}$ to $r>R_\mathrm{disk}$ may be depleted of solids. For simplicity, we do not consider this effect in our model.   
}

\revise{
To illustrate this point, we plot the cumulative pebble flux calculated with Eq.~\ref{eq: peb_flux} using $\Sigma_\mathrm{gas} (t=0)$ (solid lines) and $\Sigma_\mathrm{gas} (t=t^\prime)$ (dashed lines). We find that the two models yield similar evolution up to $t^\prime\sim\tau_\mathrm{vis}$, but the model using $\Sigma_\mathrm{gas} (t=t^\prime)$ eventually overestimates the pebble flux and violates mass conservation by exceeding the initial total solid mass. To preserve mass conservation, we adopt $\Sigma_\mathrm{gas} (t=0)$ in Eq.~\ref{eq: peb_flux}.
}

\revise{
Our pebble flux model is valid only if the pebble condensation front reaches the outer edge of the disk before the gaseous disk starts viscous evolution, i.e., when $\tau_\mathrm{peb, dep} \lesssim \tau_\mathrm{vis}$. For disks with higher viscosity or larger radial extent, a more sophisticated model of dust evolution is required to compute the pebble flux accurately.
}

\section{Mixing length parameter}\label{sec: discussion_mixing_length}

\begin{figure}
    \centering
    \includegraphics[width=1.\linewidth]{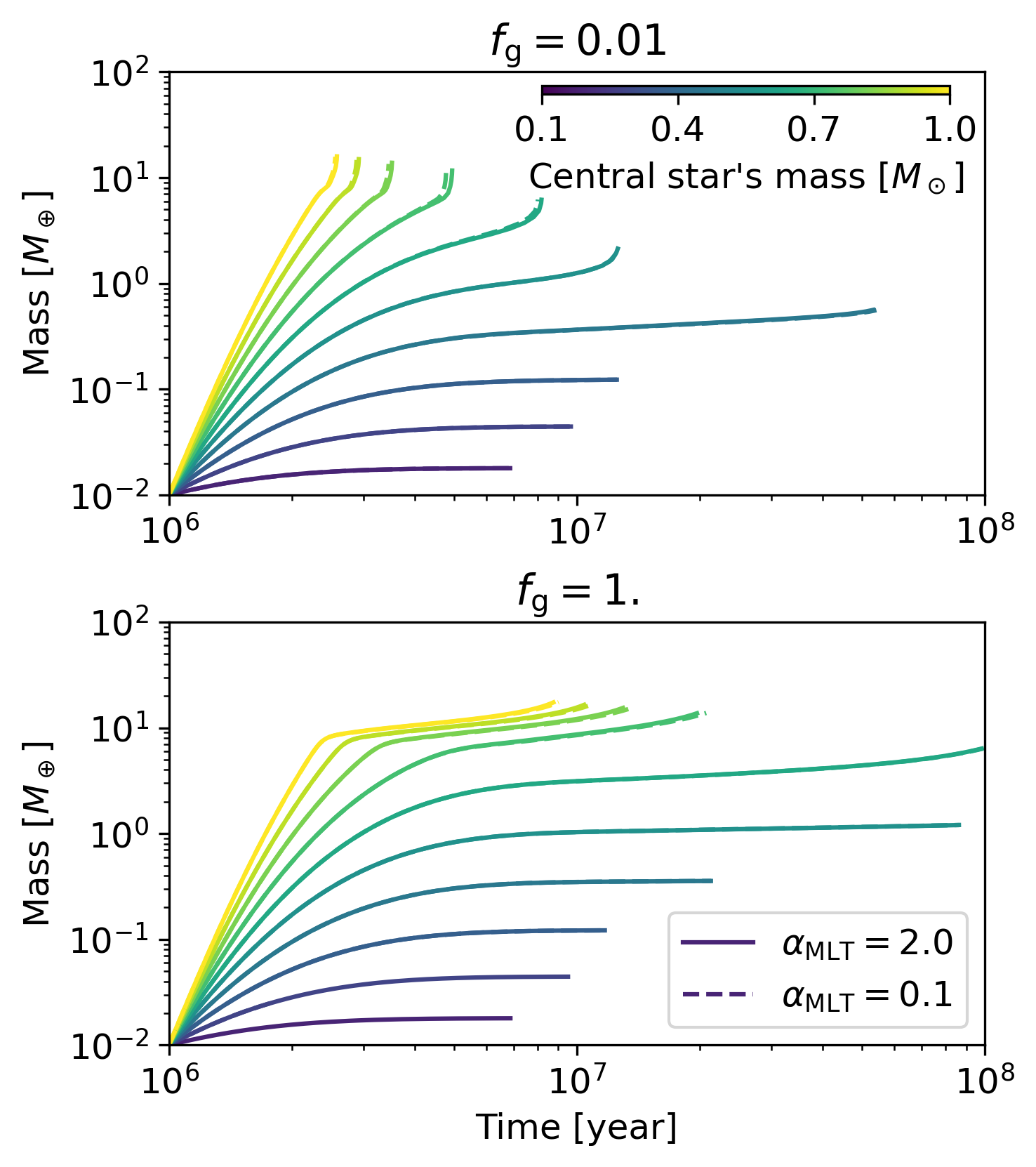}
    \caption{
    The evolution of protoplanets around different mass stars using different $\alpha_\mathrm{MLT}=2$ (solid lines) and $\alpha_\mathrm{MLT}=0.1$ (dashed lines). The top and bottom panels show the cases with different grain opacity factors $f_\mathrm{g}=0.01$ and $1$, respectively. The color corresponds to the stellar mass, as shown in the color bar.
    }
    \label{fig: psMLA}
\end{figure}

To check the effect of different mixing length parameter $\alpha_\mathrm{MLT}$, we compare the simulations with $\alpha_\mathrm{MLT}=2$ (solid lines) and $\alpha_\mathrm{MLT}=0.1$ (dashed lines) in Fig.~\ref{fig: psMLA}. We found that the difference in the obtained crossover time is less than $5\%$. We conclude that the effect of $\alpha_\mathrm{MLT}$ is negligibly smaller than that of the other parameters, such as stellar mass $M_\mathrm{s}$ and the grain opacity factor $f_\mathrm{g}$. 

\section{Supply-limited gas accretion}\label{app: gas_accretion_rate}

\begin{figure}
    \centering
    \includegraphics[width=1.\linewidth]{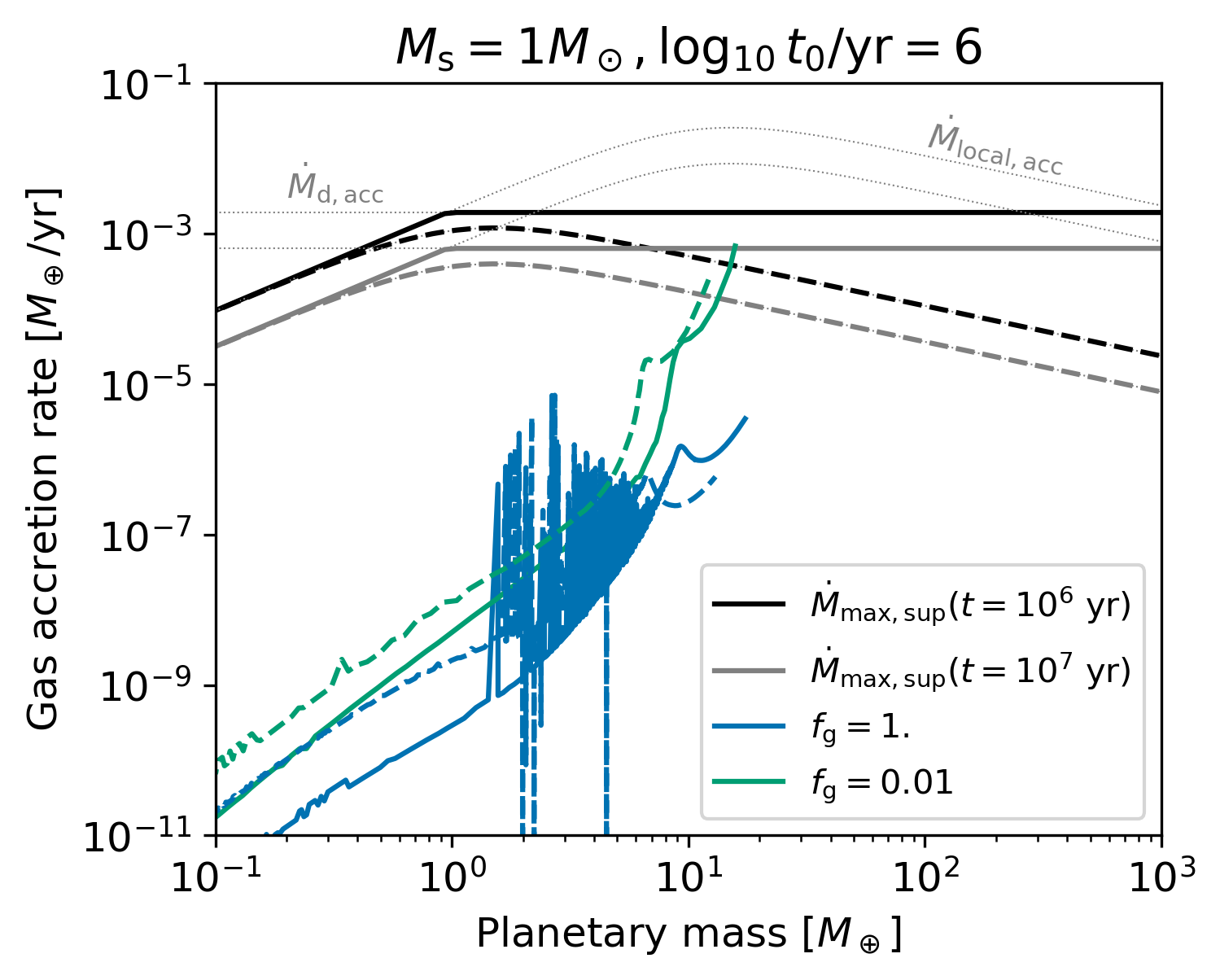}
    \caption{
    \revise{    
    Gas accretion rate obtained in our numerical simulations of $M_\mathrm{s}=1 M_\odot$ and $\log_\mathrm{10} t_0/\mathrm{yr}=6$. The blue and green lines show the cases with $f_\mathrm{g}=1$ and 0.01, respectively. The solid and dashed lines correspond to the cases with $\alpha_\mathrm{turb}=10^{-3}$ and $10^{-5}$, respectively. We also plot the maximum gas accretion rate in the supply-limited case with black ($t=10^6$ yr) and gray ($t=10^7$ yr) lines. Also, the solid and dashed lines correspond to cases with $\alpha_\mathrm{turb}=10^{-3}$ and $\alpha_\mathrm{turb}=10^{-5}$, respectively. The thin gray dotted lines 
    }
    }
    \label{fig: supply_limit}
\end{figure}

\revise{
In our simulations, the gas accretion rate is obtained by iterating the planetary radius and the gas accretion radius. The energy balance in the planetary envelope determines the gas accretion rate in this phase.
However, the gas accretion rate onto the protoplanet could be limited by the supply of local disk gas to the planetary orbital region. Once the envelope's contraction and the following gas accretion become fast enough to deplete local disk gas, the planetary envelope detaches from the protoplanetary disk and the accretion regime shifts to the supply-limited gas accretion. Here, we compare our numerical results to the maximum supply of disk gas and discuss how supply-limited gas accretion affects our results.
}

\revise{
To estimate the maximum supply of disk gas, we follow the approach developed by \citet{Tanaka+2020}. The flux of disk gas entering the protoplanet's Hill sphere is given by \citep{Tanigawa+2002}: 
\begin{align}
    \dot{M}_\mathrm{local,acc} = D \Sigma_{\rm band},  \label{eq: accretion_band}
\end{align}
with: 
\begin{align}
    D = 0.29 \left( \frac{M_{\rm p}}{M_{s}} \right)^{4/3} \left( \frac{h_\mathrm{gas}}{r} \right)^{-2} {r}^2 \Omega_{\rm K}.
\end{align}
$\Sigma_\mathrm{band}$ is the gas surface density at the gas accretion band. Protoplanets massive enough to open a gap in the gaseous disk reduce the local gas surface density and $\Sigma_\mathrm{band}$. Using the gap structure model obtained by \citet{Kanagawa+2017}, we get:
\begin{align}
    \frac{\Sigma_\mathrm{band}}{\Sigma_\mathrm{gas}} = \frac{1}{1+0.04K}, 
\end{align}
with: 
\begin{align}
    K  = \left( \frac{M_{\rm p}}{M_{s}} \right)^2 \left( \frac{h_{\rm gas}}{r} \right)^{-5} {\alpha_{\rm turb}}^{-1}.
\end{align}
Equation \ref{eq: accretion_band} shows the limit of gas disk supply through the local gas flow. In addition, the global disk accretion $\dot{M}_\mathrm{d,acc}$ limits the supply of disk gas onto protoplanets. Finally, the maximum supply of disk gas is: 
\begin{align}
    \dot{M}_\mathrm{max, sup} = \min \left( \dot{M}_\mathrm{d,acc}, \dot{M}_\mathrm{local,acc} \right). \label{eq: maximum_gas_accretion}
\end{align}
}

\revise{
Fig.~\ref{fig: supply_limit} shows the gas accretion rate obtained in our simulations and the maximum supply of disk gas $\dot{M}_\mathrm{max, sup}$. In most of the simulation phase, the gas accretion rate is slower than $\dot{M}_\mathrm{max, sup}$. In simulations with $f_\mathrm{g}=0.01$, the gas accretion rate reaches $\dot{M}_\mathrm{max, sup}$ just before  crossover mass is reached. Around crossover mass, the gas accretion timescale is much shorter than the disk's lifetime. Therefore, even when we include the effects of supply-limited gas accretion, the crossover time obtained in our simulations remains similar. 
}

\section{Heavy-element deposition in the gaseous atmosphere}\label{sec: method_Z}

We use the ablation model presented by \citet{Valletta+2020} and \citet{Mol-Lous+2024}. We use the direct deposition model, where the ablated heavy elements (from the pebbles)  are deposited into the local envelope shell. We set the composition of pebbles as 50 \% rock and 50 \% H$_2$O, and assume that only the ablated H$_2$O is deposited in the envelope. 

\citet{Valletta+2020} and \citet{Mol-Lous+2024} introduced the condensation of H$_2$O by putting the maximum metallicity of a local layer $Z_{\mathrm{max}, i}$
 (index-$i$ increases toward the center of the protoplanet). The first limit is the saturation of H$_2$O $Z_\mathrm{sat, H_2O}$. Using the saturation pressure $P_\mathrm{sat, H_2O}$, $Z_\mathrm{sat, H_2O}$ is defined as $P_\mathrm{sat, H_2O}/P_{i}$ where $P_{i}$ is the local total pressure. The second limit is the maximum water enhancement of the supercritical water $Z_\mathrm{sup, H_2O}$, which is set to 0.9, as used in \citet{Mol-Lous+2024}.

In addition, \citet{Mol-Lous+2024} introduced a smoothing factor $f_\mathrm{smooth}$ to limit the jump in the local metallicity $Z_i$. Pebbles are usually evaporated in the upper atmosphere and sink to the lower envelope. A large fraction of the heavy elements are deposited in the layer where $Z_\mathrm{sat, H_2O}$ becomes $\gtrsim0.1$. This sink of heavy elements leads to a rather large jump in $Z_i$. As a result, it is not easy for MESA to find a hydrostatic solution. To avoid a large jump in $Z_i$, \citet{Mol-Lous+2024} introduced a smoothing factor $Z_{\mathrm{smooth}, i}=Z_{\mathrm{pre},i} +f_\mathrm{smooth}$, where $Z_{pre, \mathrm{i}}$ is the local metallicity in the previous step.

The maximum metallicity of the local layer $i$ is given by: 
\begin{align}
    Z_{\mathrm{max}, i} = \min \left( Z_\mathrm{sat, H_2O}, Z_\mathrm{sup, H_2O}, Z_{\mathrm{smooth}, i} \right).
\end{align}
If the deposition of the accreted heavy elements leads to a local metallicity $Z_i$ that is higher than $Z_{\mathrm{max}, i}$, the overabundant heavy elements settle to the lower layer. The smoothing factor coefficient $f_\mathrm{smooth}$ is set to $0.2$. The smoothing factor makes the composition gradient slightly shallower than that without it. This difference in the composition gradient has a weak effect on the crossover time. We found that with the smoothing factor, the crossover times could be $5\%$ shorter at most than those obtained without it. This is because the solid core mass at the pebble isolation becomes slightly larger due to the shallower compositional gradient and the smaller envelope metallicity. This difference does not alter our main finding that the crossover time could be longer if the heavy element deposition is included in the planetary envelope. 


Pebbles evaporated in the upper atmosphere create an increasing composition gradient toward the outer envelope. MESA tries to erase this inverted composition gradient by triggering the convection under the Ledoux criterion. However, it makes the timestep too small to proceed, and numerical convergence can be difficult. To avoid this, we introduce a pre-mixing algorithm that assumes Rayleigh-Taylor instability. We cannot use the actual criterion for Rayleigh-Taylor, which compares the density of adjacent layers, because our enrichment model increases the metallicity of local shells without changing the mass there \citep[see][]{Valletta+2020}. Instead, we use this simplified criterion that checks the metallicity of adjacent layers. We mix heavy elements in and distribute them equally to the adjacent two shells if $Z_i>Z_{i+1}$.

\end{appendix}
\end{document}